\documentclass[%
 reprint,
 amsmath,amssymb,
 aps,
 pra,showkeys,
]{revtex4-2}

\usepackage{graphicx}
\usepackage{dcolumn}
\usepackage{bm}
\usepackage{hyperref}
\newtheorem{theorem}{Theorem}

\begin{document}

\preprint{APS/2309.10509}

\title{Polynomial-time Solver of Tridiagonal QUBO, QUDO and Tensor QUDO problems with Tensor Networks}

\author{Alejandro Mata Ali}
\email{alejandro.mata.ali@gmail.com}
\affiliation{i3B Ibermatica, Quantum Development Department, Paseo Mikeletegi 5, 20009 Donostia, Spain}
\author{Iñigo Perez Delgado}
\email{iperezde@ayesa.com}
\affiliation{i3B Ibermatica, Parque Tecnológico de Bizkaia, Ibaizabal Bidea, Edif. 501-A, 48160 Derio, Spain}
\author{Marina Ristol Roura}
\email{mristol@ayesa.com}
\affiliation{i3B Ibermatica, Parque Tecnológico de Bizkaia, Ibaizabal Bidea, Edif. 501-A, 48160 Derio, Spain}
\author{Aitor Moreno Fdez. de Leceta}
\email{aitormoreno@lksnext.com}
\affiliation{i3B Ibermatica, Unidad de Inteligencia Artificial, Avenida de los Huetos, Edificio Azucarera, 01010 Vitoria, Spain}

\date{\today}

\begin{abstract}
We present a quantum-inspired tensor network algorithm for solving tridiagonal Quadratic Unconstrained Binary Optimization (QUBO) problems and quadratic unconstrained discrete optimization (QUDO) problems. We also solve the more general Tensor quadratic unconstrained discrete optimization (T-QUDO) problems with one-neighbor interactions in a lineal chain. This method provides an exact and explicit equation for these problems. Our algorithms are based on the simulation of a state that undergoes imaginary time evolution and a Half partial trace. In addition, we address the degenerate case and evaluate the polynomial complexity of the algorithm, also providing a parallelized version. We implemented and tested them with other well-known classical algorithms and observed an improvement in the quality of the results. The performance of the proposed algorithms is compared with the Google OR-TOOLS and dimod solvers, improving their results.
\end{abstract}

\keywords{Tensor networks, Quantum-inspired, Combinatorial optimization, Quantum computing}
\maketitle


\section{Introduction}
Quadratic Unconstrained Binary Optimization (QUBO) \cite{QUBO} problems are a type of combinatorial optimization problem that lies at the intersection of quantum computing and optimization theory. These problems are characterized by their ability to represent a wide variety of complex challenges and their use in industrial and applied fields such as logistics \cite{Logist, Logist2}, engineering \cite{Engin}, physics \cite{QUBO}, biology \cite{Biol}, and economics \cite{Finances}. In a QUBO problem, one seeks to find an assignment of binary values (0 or 1) to a set of decision variables such that the quadratic function of these variables is minimized. This quadratic function, also known as a cost or energy function, can represent constraints and objectives of a specific problem. Quadratic Unconstrained Discrete Optimization (QUDO) problems are a generalization of the QUBO problems, allowing the variables to take a larger number of integer values and can be translated to QUBO problems with a logarithmic amount of binary QUBO variables for each discrete QUDO variable. In order to represent more complex problems, other formalisms have been developed. The most well-known generalization is the Higher-Order Unconstrained Binary Optimization (HOBO) problem, which has a cost function that involves products of more than two binary variables. Another novel approach is the Tensor Quadratic Unconstrained Discrete Optimization (T-QUDO) problem~\cite{t_qudo}, which uses tensor positions instead of products to represent more naturally certain types of problem. Both problems have a degree of complexity too high to be efficiently solved classically \cite{Complex}, except in certain cases \cite{Polynomial,Zero-One}. As a result, they are often tackled using approximate or heuristic methods, such as genetic algorithms \cite{GPU}, linear programming \cite{1D_Ground} or digital annealing \cite{Digital}.

QUBO problems are particularly interesting and relevant in the era of quantum computing because of their ability to take advantage of the properties of quantum computing. This is due to the equivalence between the QUBO problems and the Ising model, allowing algorithms such as the Quantum Approximate Optimization Algorithm (QAOA)\cite{qaoa} to solve them. It has been applied in digital quantum computing \cite{Grover}, quantum annealing \cite{Hobbit}, and hybrid algorithms \cite{Logist, Logist2}.  QUDO problems are also compatible with quantum computing \cite{QUDO_QAOA}, since it is possible to make use of qudits or groups of qubits that simulate to be one qudit, or transform the QUDO into a QUBO problem.

However, due to the current state of quantum hardware, noisy and with small capacity~\cite{nisq}, and its availability, the field of quantum-inspired technologies has gained importance. These classical technologies involve taking advantage of certain quantum properties to accelerate calculations. One of the most important of these is tensor networks \cite{Tensor_Network,orus_tn}, which use algebraic mathematics of quantum systems to simulate them classically and extract certain properties of the simulated systems. They can implement operations that would not be possible in quantum systems, such as forced post-selection or the application of non-unitary operators. This technology makes possible the compression of information, both for classical and quantum systems, and has been applied to machine learning~\cite{TNN}, large language model~\cite{compactifai} compression, and quantum algorithm simulation~\cite{Simul_tn}.

For solving 1D QUBO problems, there are algorithms in tensor networks, such as \cite{1D_Ground}. There are also other techniques for more general combinatorial optimization problems~\cite{TTOPT,Optimiz}. However, these have a high computational complexity that may make them not optimal for large instances.

In this paper, we will explore an efficient tensor network algorithm to solve general QUBO, QUDO, and Tensor QUDO problems with one-neighbor interactions in a lineal chain. We call it the \textit{Lineal Chain Quadratic Problem Tensor Network Solver} (LCQPTNS). The connection of these problems with Ising models could lead to applications in quantum ground-state simulation with the same formalism. It will consist of a sequential tensor network contraction algorithm that simulates an imaginary time evolution and a partial trace. We will also analyze its applicability to degenerate cases and see its computational complexity. Our main contributions are
\begin{itemize}
    \item Provide a novel methodology for solving 1D combinatorial problems.
    \item Provide a novel algorithm for solving general QUBO, QUDO and Tensor QUDO problems with one-neighbor interactions in a lineal chain. To the best of our knowledge, this is the first algorithm to address this exact problem directly.
    \item Provide the first parallelized algorithm for MeLoCoToN problem resolution applications.
    \item Provide an explicit and exact equation that returns the solution for the three problems.
    \item Analyze the performance of the algorithm, theoretically and empirically, and compare it with other algorithms.
    \item Provide a Python implementation of the algorithms.
\end{itemize}

This work is structured as follows. First, we describe the problems we want to solve in Sec.~\ref{sec: description} and we present a brief background on the state-of-the-art in solving them. Then, we introduce the three novel algorithms in Sec.~\ref{sec:tn}. Finally, we performed several experiments in Sec.~\ref{sec:experiments} to analyze its performance.

All code is publicly available on the GitHub repository \href{https://github.com/DOKOS-TAYOS/Lineal_chain_QUBO_QUDO_TensorQUDO_Solver_with_Tensor_Networks}{https://github.com/DOKOS-TAYOS/Lineal\_chain\_QUBO\_QUDO\_TensorQUDO}

\section{Description of the problem}\label{sec: description}
A general QUBO problem can be expressed by a quadratic cost function to minimize using a vector $\Vec{x}$ of $N$ binary components. That is, we look for an optimal $\Vec{x}_{\text{opt}}$ such that
\begin{equation}
    \begin{gathered}
            \Vec{x}_{\text{opt}} = \arg \min_{\Vec{x}} C(\Vec{x}),\\
            \quad x_i \in \{0,1\}, \quad i \in [0,N-1]\\
    C(\Vec{x})=\sum_{\substack{i,j=0 \\ i\leq j}}^{N-1} w_{ij}x_i x_j,
    \end{gathered}
\end{equation}
where $w_{ij}$ are the elements of the weight matrix $w$ of the problem and $C(\cdot)$ is the cost function of the problem. The diagonal elements $w_{ii}$ are the local terms, and the non-diagonal elements $w_{ij}$ are the interaction terms.

In a QUDO case, the problem is analogous, changing that the components of $\vec{x}$ will be integers in a certain range, not only $0$ or $1$. This is
\begin{equation}
\begin{gathered}
    \Vec{x}_{\text{opt}} = \arg \min_{\Vec{x}} C(\Vec{x})\\
    x_i \in \{0,1,\dots, D_i-1\}, \quad i \in [0,N-1]\\
    C(\Vec{x})=\sum_{\substack{i,j=0 \\ i\leq j}}^{N-1} w_{i,j}x_i x_j + \sum_{i=0}^{N-1} d_{i}x_i,
\end{gathered}
\end{equation}
being $d_i$ the elements of the cost vector $\vec{d}$ of the problem and $D_i$ the number of possible values of the $i$-th variable. QUDO problems can be transformed into QUBO problems transforming $x_i$ variables into a set of $s_{i,k}$ variables
\begin{equation}
    x_i = \sum_{k=0}^{ \log_2 D_i - 1} 2^k s_{i,k},
\end{equation}
if $\log_2 D_i$ is an integer.

In a Tensor QUDO case, the cost function of the problem now is formulated as the sum of the elements of a tensor, selected by the solution vector. This is
\begin{equation}
\begin{gathered}
    \Vec{x}_{\text{opt}} = \arg \min_{\Vec{x}} C(\Vec{x}),\\
    x_i \in \{0,1,\dots, D_i-1\}, \quad i \in [0,N-1]\\
    C(\Vec{x})=\sum_{\substack{i,j=0 \\ i\leq j}}^{N-1} w_{i,j,x_i, x_j}
\end{gathered}
\end{equation}
where $w_{i,j,x_i, x_j}$ are the elements of the weight tensor $w$ of the problem, which depends on the value of the variables and their positions in the solution. Both QUBO and QUDO problems are particular cases of Tensor QUDO problems.

A special case of interest is the case of nearest-neighbor interaction in a linear chain, which can be understood as the Ising model in one dimension. In this problem, each variable interacts only with the one it has before and with the one it has just after. Therefore, our QUBO and QUDO problems cost function simplifies to
\begin{equation}
    C(\Vec{x})= \sum_{i=0}^{N-1} (w_{i,i}x_i^2+d_ix_i)+\sum_{i=0}^{N-2} w_{i,i+1}x_i x_{i+1},
\end{equation}
which implies that $w$ is a tridiagonal matrix
\begin{equation}
    w = \begin{bmatrix}
    w_{11} & w_{12} & 0 & 0 & 0 \\
    w_{21} & w_{22} & w_{23} & 0 & 0 \\
    0 & w_{32} & w_{33} & w_{34} & 0 \\
    0 & 0 & w_{43} & w_{44} & w_{45} \\
    0 & 0 & 0 & w_{54} & w_{55}
    \end{bmatrix}.
\end{equation}

In the Tensor QUDO case, the analogous problem would be
\begin{equation}
    C(\Vec{x})=\sum_{i=0}^{N-1} w_{i,i,x_i, x_i}+\sum_{i=0}^{N-2} w_{i,i+1,x_i, x_{i+1}}
\end{equation}
which could be considered a kind of shortest path problem without restrictions, where the distances change with every time step.

There exist several algorithms to solve QUBO problems, from classical ones, such as Spiking Neural Networks~\cite{spiking_qubo,Zero-One,GPU,1D_Ground}, to quantum computing ones, such as QAOA~\cite{qaoa}. However, in the state-of-the-art, there are no prior algorithms that address the lineal chain problem exactly and efficiently. The most efficient algorithm for this 1D problem is presented in~\cite{1D_Ground}, but even in the most simple case, it runs as $O(n^{16})$. And this algorithm also scales exponentially with $D_i$ for the QUDO and Tensor QUDO problems, so it is not efficient in these cases. Taking into account that the general QUBO problem is an NP-Hard problem, all known exact algorithms to solve it have exponential complexity.

In heuristic algorithms to approach this type of 1D problem, the Density Matrix Renormalization Group (DMRG)~\cite{dmrg}  is the most well-known algorithm. However, this method is only effective if the bond dimension does not scale exponentially. Its usual complexity in QUDO cases would be $O(ND^3\chi^3)$ for each sweep, being $D$ the typical dimension of the variables and $\chi$ the required bond dimension, and could be extremely low for large systems. Additionally, this algorithm is not designed for Tensor QUDO problems.

\section{Tensor Network algorithms}\label{sec:tn}
In this section we introduce the new tensor network algorithms for solving QUBO, QUDO and Tensor QUDO problems in a lineal chain with one-neighbor interaction efficiently. First, in Ssec.~\ref{ssec: tridiagonal QUBO} we introduce the QUBO algorithm, as a first and simpler version. Secondly, in Ssec.~\ref{ssec: QUDO} we introduce how to generalize it for QUDO problems. Third, in Ssec.~\ref{ssec: T-QUDO} we generalize it for Tensor QUDO problems. Then, we discuss how to optimize the tracing method in Ssec.~\ref{ssec:optimization} and to address the degenerate case in Ssec.~\ref{ssec:degenerate}. Finally, in Ssec.~\ref{ssec:complexity} we analyze the computational complexity of the algorithms execution.

\subsection{Tridiagonal QUBO tensor network solver}\label{ssec: tridiagonal QUBO}
First we will create a solver for the tridiagonal QUBO problem and then we will see how to generalize it to the QUDO problem. We use a modified version of the method \cite{Optimiz}, and it is the first algorithm to use the MeLoCoToN formalism~\cite{melocoton}. Our algorithm can be summarized in the following theorem.

\begin{theorem}
Given a QUBO problem described by a tridiagonal $N\times N$ weight matrix $w$ and $N$ indeterminate binary variables, we can determine an approximate optimal solution in $O(N)$ time, which is the optimal in the limit as $\tau\rightarrow\infty$.
\end{theorem}

\begin{figure}[h]
    \centering
    \includegraphics[width=0.45\textwidth]{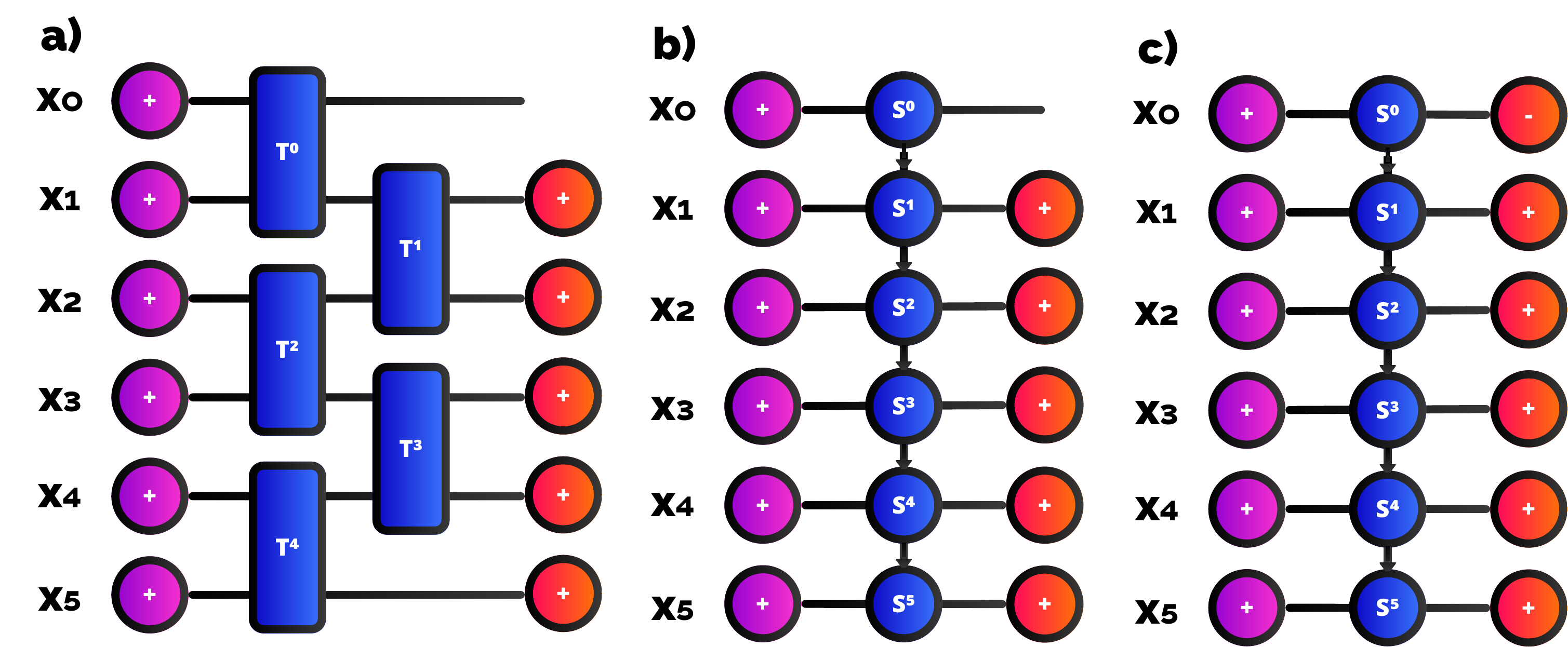}
    \caption{Tensor networks solving the tridiagonal QUBO problem and the one-neighbor QUDO problem. a) Bilocal version, b) Matrix Product Operator (MPO) version, c) Matrix Product Operator (MPO) version for explicit equation.}
    \label{fig:QUBO_TN}
\end{figure}

For this, we will explain the basics of the algorithm with the tensor network of Fig. \ref{fig:QUBO_TN} a, which mimics the shape of a quantum circuit, with each horizontal line being the timeline of a qubit and the vertical lines controlling operations between them. The `+' nodes represent a qubit in uniform superposition $(1,1)$ and the $T$ nodes represent the imaginary time evolution that depends on the state of the two neighboring qubits.

The `+' nodes represent the state of each variable $x_i$. It can be visualized as the initialization of a quantum circuit. By performing the tensor product with all these tensors, each in uniform superposition, we will have uniform superposition of all $2^N$ combinations. That is, a vector of $2^N$ ones.

The goal is for our tensor layer $T$ to encode the state in our tensor network as
\begin{equation}
    |\psi\rangle = \sum_{\vec{x}} e^{-\tau C(\vec{x})} |\vec{x}\rangle,
\end{equation}
so that the combination $\vec{x}$ with the lowest cost has an exponentially larger amplitude than the other combinations. In tensor terms, the tensor represented by this tensor network is
\begin{equation}
    \mathcal{T}_{\vec{x}} = e^{-\tau C(\vec{x})}.
\end{equation}

\begin{figure}[h]
    \centering
    \includegraphics[width=0.45\textwidth]{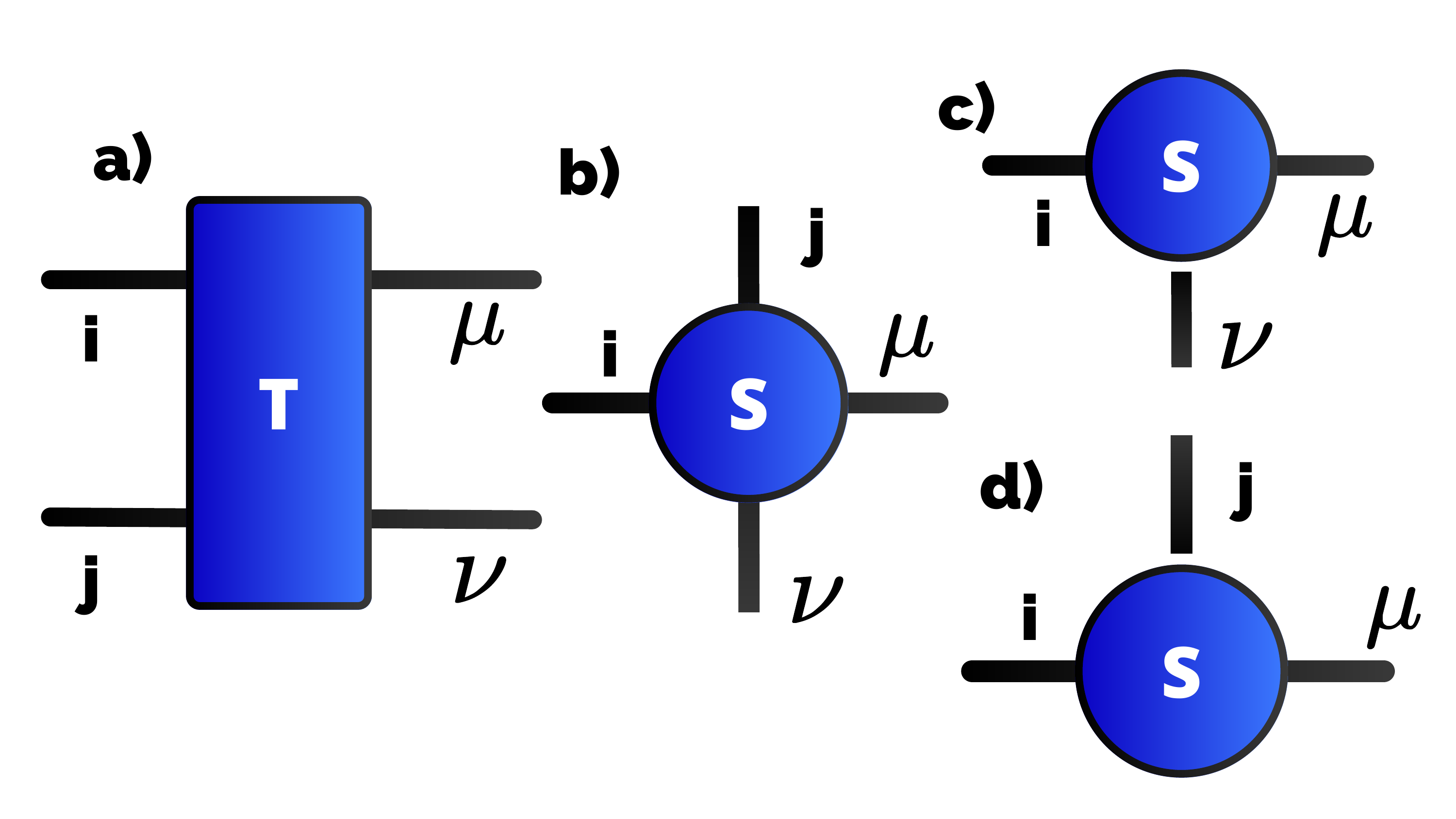}
    \caption{Tensor indices. a) $T$ tensor, b) First $S$ tensor, c) Intermediate $S$ tensor, d) Final $S$ tensor.}
    \label{fig:QUBO_Tensors}
\end{figure}

The $T$ tensors will be tensors with four 2-dimensional indices $i,j,\mu,\nu$ (Fig. \ref{fig:QUBO_Tensors}) whose non-zero elements correspond to cases where $\mu=i$ and $\nu=j$. This means that the state of the first variable enters at index $i$ and exits at index $\mu$ and the state of the second variable enters at index $j$ and exits at index $\nu$. Using the sparse logical notation introduced in~\cite{melocoton}, the non-zero elements are
\begin{equation}
\begin{gathered}
    T^n_{2\times 2\times 2\times 2},\\
    \mu=i, \nu=j,\\
    T^n_{ij\mu\nu} = e^{-\tau (w_{n,n+1} ij + w_{n,n}i)},
\end{gathered}
\end{equation}

\begin{equation}
\begin{gathered}
    T^{N-2}_{2\times 2\times 2\times 2},\\
    \mu=i, \nu=j,\\
    T^{N-2}_{ij\mu\nu} = e^{-\tau (w_{N-2,N-1} ij + w_{N-2,N-2}i+w_{N-1,N-1}j)},
\end{gathered}
\end{equation}
where $\tau$ is a decay hyperparameter, related to the evolution in imaginary time, and $T^n$ is the $T$-tensor which connects the $n$-th and $n+1$-th variables.

Since the resulting tensor has $2^N$ components, we cannot simply look at which component has the largest amplitude. We will extract its information in a more efficient way. We assume that the amplitude of the lowest-cost combination is sufficiently larger than the amplitudes of the other combinations. This is a reasonable assumption because if we increase $\tau$, the combinations will change their amplitudes exponentially differently. In the infinite limit, only the relative amplitude of the optimal solution will remain. From this tensor we will obtain the value of $x_0$ associated with the optimal combination. If there is a combination with a sufficiently higher amplitude, adding the amplitudes of all combinations with $x_0 = 0$ and with $x_0=1$ separately, the main contribution will be the one with the highest amplitude. We will call this operation \textit{Half Partial Trace} with $x_0$ free, which returns a vector $P^{x_0}$ with components.
\begin{equation}
    P^{x_0}_i=\sum_{\substack{\vec{x}\\ x_0=i}} e^{-\tau C(\vec{x})}.
\end{equation}
This partial trace is made by connecting a `+' tensor to each variable at the output of the $T$ tensor layer, except for the $x_0$ variable. Taking the limit of large $\tau$, assuming no degeneracy,
\begin{equation}
    \lim_{\tau\rightarrow \infty} \frac{P^{x_0}_i}{\sum_iP^{x_0}_i}=
    \lim_{\tau\rightarrow \infty}\frac{\sum_{\substack{\vec{x}\\ x_0=i}}e^{-\tau C(\vec{x})}}
    {\sum_i\sum_{\substack{\vec{x}\\ x_0=i}}e^{-\tau C(\vec{x})}}. 
    = \delta_{i,(x_{opt,0})}
\end{equation}

Out of the limit, if the combination with the highest amplitude has $x_0=0$, then $P^{x_0}_0>P^{x_0}_1$, and in the opposite case $P^{x_0}_0<P^{x_0}_1$. This also means that, if we multiply the resulting vector by a \textit{minus vector} $(-1,1)$, if the correct value is $x_0=0$, so $P^{x_0}=(1,0)$, then the contraction of the tensor network is $-1$, and if it is $x_0=1$, then it results in $+1$. So, the correct value can be expressed as
\begin{equation}
    x_0 = H(\Omega_0(\tau))
\end{equation}
namely $H(\cdot)$ the Heaviside step function and $\Omega_0(\tau)$ the tensor network described with the minus vector in the $0$-th position for a $\tau$ value. This is represented in Fig.~\ref{fig:QUBO_TN} c for six variables.

We can optimize the contraction of the tensor network by defining it as shown in Fig. \ref{fig:QUBO_TN} b, where we replace the $T$-tensors by a Matrix Product Operator (MPO) layer of $S$-tensors. These tensors perform exactly the same function, sending signals up and down through their vertical bond indices. All of their indices are of dimension 2. In the figure, the arrows indicate the flow of information through the nodes of the layer. These will tell adjacent $S$ tensors about their associated variable state and, depending on the signal they receive from the previous $S$ tensor and their own variable, apply a certain imaginary time evolution. This tensor network is much easier and more straightforward to contract. We call $S^n$ the $S$-tensor connected with the variable $x^n$.

Therefore, the non-zero elements of the $S$-tensors will be those where $\mu=i$ for $S^{N-1}$ and $\mu=\nu=i$ for the others. Their values will be
\begin{equation}
\begin{gathered}
    S^0_{2\times 2\times 2},\\
    \mu=\nu=i,\\
    S^0_{i\mu\nu} = e^{-\tau (w_{0,0}i)},
\end{gathered}
\end{equation}
\begin{equation}
\begin{gathered}
    S^n_{2\times 2\times 2\times 2},\\
    \mu=\nu=i,\\
    S^n_{ij\mu\nu} = e^{-\tau (w_{n-1,n}ij+w_{n,n}i)},
\end{gathered}
\end{equation}
\begin{equation}
\begin{gathered}
    S^{N-1}_{2\times 2\times 2},\\
    \mu=i,\\
    S^{N-1}_{ij\mu} = e^{-\tau (w_{N-2,N-1}ij+w_{N-1,N-1}i)}.
\end{gathered}
\end{equation}

With the corresponding tensor network, we first determine the variable $x_0$, then follow these steps to get the other components:
\begin{enumerate}
    \item Perform the algorithm for the $x_0$ component.
    \item For $n\in[1, N-2]$:
        \begin{itemize}
            \item Carry out the same algorithm, but eliminating the previous $n-1$ variables. Change the new tensor $S^0$ so that its components match those of the tensor $S^{1}$ from the previous step, with its index $j$ set to $x_{n-1}$. This means
            \begin{equation}
                S^0_{i\mu\nu} = S^{1, \text{previous}}_{i,x_{n-1},\mu,\nu}.
            \end{equation}
            The result of the algorithm will be $x_{n}$.
        \end{itemize}
    \item For $x_{N-1}$, we use the classical comparison:
    \begin{equation}
        x_{N-1} = H(-(w_{N-1, N-1}+w_{N-2,N-1}x_{N-2})).
    \end{equation}
\end{enumerate}
The progressive reduction of the represented variables is due to the fact that if we already know the solution, we can consider the rest of the unknown system and introduce the known information into the system. The redefinition of the tensor $S^0$ in each step allows us to consider the result of the previous step in obtaining the costs of the variable pair to be determined and the previous one.

To obtain the exact explicit equation, we will omit the variables reduction to simplify the explanation. If we have a tensor network of layers `+' and $S$, we can extract the $n$-th variable connecting the `+' nodes for all the variables, but the $n$-th one, which will be connected to a minus vector. We call the resulting tensor network $\Omega_n(\tau)$. We know that in the limit this algorithm is exact if we have no degeneracy, so we can express
\begin{equation}
    x_n =\lim_{\tau\rightarrow\infty} H(\Omega_n(\tau)).
\end{equation}
This equation is exact because all the arguments previously shown and explicit because the solution does not depend on the own solution we want. So, the Heavyside step function of this tensor network is the equation for the solution of the problem. Then, this is an exact and explicit equation to solve the tridiagonal QUBO problem.

\subsection{Tridiagonal QUDO problem}\label{ssec: QUDO}
In the QUDO problem, we can use the same tensor network structure. We will only have to change our binary variable formalism to a discrete variable formalism and allow the indices of the tensor network in Fig. \ref{fig:QUBO_TN} to have the dimension required for each variable. From now on, the variable $x_i$ will have $D_i$ possible values. Our algorithm can be summarized in the following theorem.

\begin{theorem}
Given a QUDO problem described by a tridiagonal $N\times N$ weight matrix $w$ and $N$ indeterminate binary variables of $D$ possible values, we can determine an approximate optimal solution in $O(ND^2)$ time, which is optimal in the limit as $\tau\rightarrow\infty$.
\end{theorem}

Each `+' tensor will now have $D_i$ components, depending on the variable they represent, all only with ones. The same is true for those we use to make the partial trace. The construction is the same, but the new $S$-tensor definitions are
\begin{equation}
\begin{gathered}
    S^0_{D_0\times D_0\times D_0},\\
    \mu=\nu=i,\\
    S^0_{i\mu\nu} = e^{-\tau (w_{0,0}i^2+d_0 i)},
\end{gathered}
\end{equation}
\begin{equation}
\begin{gathered}
    S^n_{D_n\times D_{n-1}\times D_n\times D_n},\\
    \mu=\nu=i,\\
    S^n_{ij\mu\nu} = e^{-\tau (w_{n-1,n}ij+w_{n,n}i^2+d_n i)},
\end{gathered}
\end{equation}
\begin{equation}
\begin{gathered}
    S^{N-1}_{D_{N-1}\times D_{N-2}\times D_{N-1}},\\
    \mu=i,\\
    S^{N-1}_{ij\mu} = e^{-\tau (w_{N-2,N-1}ij+w_{N-1,N-1}i^2+d_{N-1} i)}.
\end{gathered}
\end{equation}
As in the QUBO case, the $S$-tensors receive the state of the adjacent variable through the index $j$ and send theirs through the index $\nu$.

The rest of the algorithm is the same until the value extraction. If we contract this tensor network analogously to the QUBO case, we obtain a vector $P^{x_0}$ of dimension $D_0$. From this, we can extract the optimal value of $x_0$ by looking for the largest component. That is, if the vector obtained by the tensor network were $(2, 4, 27, 2, 0, 1)$, the correct value for $x_0$ would be $2$. To obtain the other variables, we will perform exactly the same process that we have explained for the QUBO case. We will change the last step to a comparison that gives us the value of $x_{N-1}$, which will result in a lower cost based on the value of $x_{N-2}$ we already have.

To obtain the explicit equation, in the index of the variable we want to determine, we cannot simply connect a minus vector because the dimensions do not fit. However, we can binarize the value of $x_n$ into a set of binary variables $x_{n,m}$. This means that instead of connecting a minus vector, we connect a \textit{bit-selector vector} $B$. To exemplify this vector, for a dimension four case, it is $(-1,1,-1,1)$ for the first bit determination and $(-1,-1,1,1)$ for the second one. In general, to determine the $m$ bit having a dimension $D$, the $B^{D,m}$ vector $i$-th component is defined as
\begin{equation}
    B^{D,m}_i= (-1)^{b(i)_m+1},
\end{equation}
being $b(i)$ the binary vector of the binarization of $i$.

This means that to determine the $m$-th bit of the $x_n$ variable, the equation is the following.
\begin{equation}
    x_{n,m} = \lim_{\tau\rightarrow\infty} H(\Omega_{n,m}(\tau)),
\end{equation}
being $\Omega_{n,m}(\tau)$ the tensor network defined for the problem, Half Partial traced for all the variables, but $n$-th one, which is connected with $B^{D_n,m}$. This is an exact and explicit equation to solve the tridiagonal QUDO problem.

\subsection{One-neighbor Tensor QUDO problem}\label{ssec: T-QUDO}
In this case, we are going to solve the lineal chain one-neighbor Tensor QUDO problem using an extended version of the tridiagonal QUDO algorithm. Here, we only need to change the value of the elements of the $S$-tensors.

Our algorithm can be summarized in the following theorem.
\begin{theorem}
    Given a Tensor QUDO problem described by a lineal chain one-neighbor $N\times N\times D \times D$ weight tensor $w$ and $N$ indeterminate variables with $D$ possible values, we can determine an approximate optimal solution in $O(ND^2)$ time, which is optimal in the limit as $\tau\rightarrow\infty$.
\end{theorem}

Once again, the elements of the $S$-tensors are non-zero when $\mu=i$ for $S^{N-1}$, and $\mu=\nu=i$ for the others. They are
\begin{equation}
\begin{gathered}
    S^0_{D_0\times D_0\times D_0},\\
    \mu=\nu=i,\\
    S^0_{i\mu\nu} = e^{-\tau w_{0,0,i,i}},
\end{gathered}
\end{equation}
\begin{equation}
\begin{gathered}
    S^n_{D_n\times D_{n-1}\times D_n\times D_n},\\
    \mu=\nu=i,\\
    S^n_{ij\mu\nu} = e^{-\tau (w_{n-1,n,i,j}+w_{n,n,i,i})},
\end{gathered}
\end{equation}
\begin{equation}
\begin{gathered}
    S^{N-1}_{D_{N-1}\times D_{N-2}\times D_{N-1}},\\
    \mu=i,\\
    S^{N-1}_{ij\mu} = e^{-\tau (w_{N-2,N-1,i,j}+w_{N-1,N-1,i,i})}.
\end{gathered}
\end{equation}

The rest of the algorithm is the same as in the QUDO problem, and the exact and explicit equation is constructed in the same way.

\subsection{Tracing optimization}\label{ssec:optimization}
One of the biggest problems we may encounter is choosing a value of $\tau$ large enough to distinguish the optimal combination but not so large that the amplitudes go to zero. For this reason, in practice, it is advisable to rescale the $w$ matrix before starting so that the minimum cost scale does not vary too much from problem to problem. In this way, we can leave $\tau$ constant. A good practice to ensure this may be that the maximum cost we can obtain is on the order of 1 and the minimum on the order of -1 or 0, depending on the characteristics of the problem.

In addition, an effective way to modify the general algorithm is to initialize in a superposition state with complex phases between the base combinations instead of initializing with a superposition with the same phase. This allows us that, when we perform the Half Partial Trace, instead of summing all amplitudes in the same direction, they will be summed as 2-dimensional vectors. This facilitates the suboptimal states to be damped against each other, allowing the maximum to be seen better.

We add these phases by initializing the `+' tensors in $(e^{2\pi i\cdot 0 \frac{1}{D_n}},e^{2\pi i\cdot 1 \frac{1}{D_n}},e^{2\pi i\cdot 2 \frac{1}{D_n}},\dots,e^{2\pi i\cdot(D_n-1)\frac{1}{D_n}})$ instead of in $(1,1, \dots,1)$. Thus, each possible combination has its own associated phase. We call it the \textit{Humbucker} method, inspired in the noise cancellation for guitar coils. To improve performance, we can add a small random factor to each combination phase.

In this case, we only need to change the definition of $P^{x_0}$ to
\begin{equation}
    P^{x_0}_i=\left\vert\left\vert\sum_{\substack{\vec{x}\\ x_0=i}} e^{i\gamma_{\vec{x}}} e^{-\tau C(\vec{x})}\right\vert\right\vert,
\end{equation}
being $\gamma_{\vec{x}}$ the phase of the $\vec{x}$ state.

Another problem that can arise is to have an $N$ such that when adding the $D^N$ damped states, we exceed the precision of the numbers we are using, obtaining divergences in the trace vector. Related to this problem, we have the possibility that by multiplying so many nodes, we obtain all 0 values due to lack of precision in our numbers. To deal with this, we can choose to divide the components of the $n$ initialization `+' nodes of step $n$ so that we try to make the sum of the components of the final trace vector near $D_n$. That is, we will try to have a 1-norm
\begin{equation}
    \sigma_n = \sum_i P^{x_0}_i = D_n.
\end{equation}
We choose $D_n$ as the target value to account for the possibility of having large differences in $D$ between the problems and the variables. To do so, we will make each initialization node `+' at step $n$ multiply its components by
\begin{equation}
    \left(\frac{1}{D_n}\right)^\frac{1}{N-n} \xi_n = \left(\frac{1}{D_n}\right)^\frac{1}{N-n} \xi_{n-1}^{\frac{N-n+1}{N-n}}  \left(\frac{1}{\sigma_{n-1}}\right)^\frac{1}{N-n}
\end{equation}
where $\xi_n$ is the factor that accounts for how much $\left(\frac{1}{\sigma_{n-1}}\right)^\frac{1}{N-n}$ would be if we did not apply this normalization, with $\xi_0=1$.

By doing this, we will be normalizing the states so that the state after the time evolution will have a 1-norm closer to a controlled value, avoiding divergences. Another possibility is to implement the normalization algorithm presented in~\cite{Initi}.

\subsection{Degenerate case}\label{ssec:degenerate}
Until now, we have considered a non-degenerate case. However, in the degenerate case, where we have more than one optimal combination, we have more peaks of similar amplitude. They are not exactly equal, because of the contribution of residual states. In the phased method, we can have the situation where the two peaks have opposite phases, so that they cancel out and we cannot see the optimum. For this reason, we recommend that the phased version should not be used in case of a possible degeneration of the problem.

Our method without phases allows us to avoid the degeneracy problem, because if at any step of the process we have two peaks with two different values of $x_n$, we will choose one of them and the rest of the combination we obtain will be the one associated exactly with the $x_n$ we have chosen, avoiding the degeneracy problem. In addition, if we want the other states of the degeneracy, we can do the same process again, but choosing the other high-amplitude component instead.

\subsection{Complexity analysis}\label{ssec:complexity}
\begin{figure}[ht]
    \centering
    \includegraphics[width=0.45\textwidth]{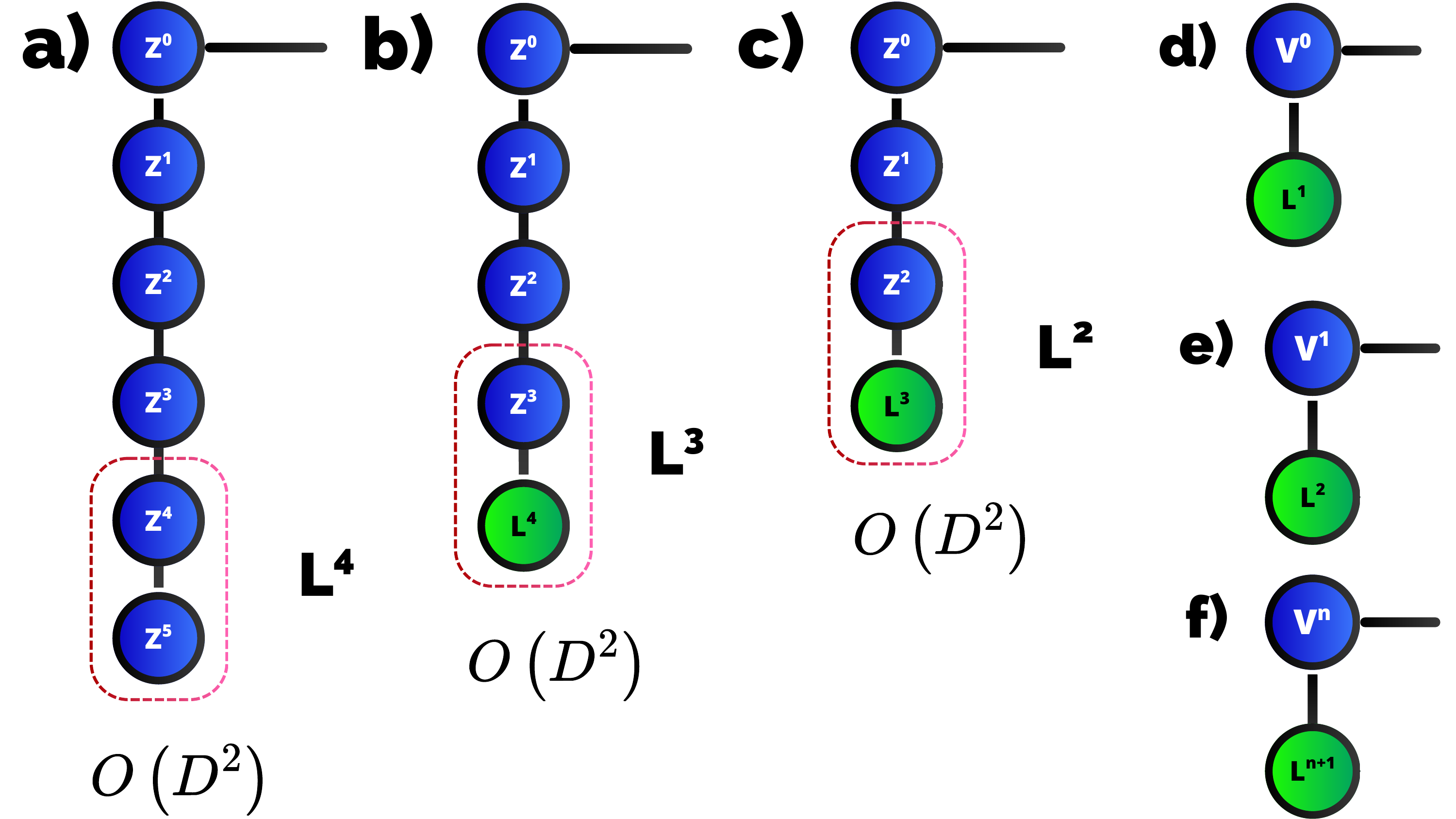}
    \caption{Contraction scheme for the efficient tensor network. We have to repeat this process $N-1$ times to obtain a component. a) Partial trace of the last variable. b) Partial trace of the penultimate variable. c) Contraction of the last variable with the penultimate one. d)}
    \label{fig:Contraction}
\end{figure}

Before analyzing its computational complexity, we have to emphasize that the tensor network can be further optimized. We take into account the fact that the contraction of the `+' tensors with the $S$ tensors only implies that the resulting tensors will be the same as the $S$ tensors that created them, but by simply eliminating the $i$ index. For the traced variables, it also means removing the $\mu$ index, but in the last one, which implies the sum of the values for all the $i$ values. So, the optimal tensor network would be exactly the same as in Fig. \ref{fig:QUBO_TN} b, but eliminating the first and last layers of the `+' tensors and their associated indices in the $S$ tensors. This tensor network is the one in Fig. \ref{fig:Contraction} a.

The definition of $Z$-tensors in the Tensor QUDO case is
\begin{equation}
\begin{gathered}
    Z^0_{D_0\times D_0},\\
    \nu=i,\\
    Z^0_{i\nu} = e^{-\tau w_{0,0,i,i}},
\end{gathered}
\end{equation}
\begin{equation}
\begin{gathered}
    Z^n_{D_{n-1}\times D_n},\\
    Z^n_{ij} = e^{-\tau (w_{n-1,n,i,j}+w_{n,n,i,i})},
\end{gathered}
\end{equation}
\begin{equation}
\begin{gathered}
    Z^{N-1}_{D_{N-1}},\\
    Z^{N-1}_{j} = \sum_{i=0}^{D_{N-1}}e^{-\tau (w_{N-2,N-1,i,j}+w_{N-1,N-1,i,i})},
\end{gathered}
\end{equation}
and analogous for the QUBO and QUDO problems. The complexity of generating these tensors is $O(D^2)$ for each one, the same for their space complexity. Taking into account that they are $N$ different, we have a computational and space complexity of $O(ND^2)$ for the creation and storage of tensors. However, if we create them just before using them, we can store only one matrix at each step, so the complexity would be $O(D^2)$.

The computational complexity of contracting each of the tensor networks for a QUDO problem with $N$ variables that can take $D$ values is $O(N D^2)$, because it multiplies $N$ matrices $D\times D$ by a vector. This is due to having to apply the contraction scheme in Fig. \ref{fig:Contraction}. We have to repeat it $N-1$ times to determine the $N$ variables, so the total computational complexity of the algorithm is $O(N^2D^2)$, the space complexity remains in $O(D^2)$, and with solution storage it is $O(N+D^2)$. However, storage of the $w$ tensor of Tensor QUDO requires $O(ND^2)$ space complexity, so that is the space complexity in this case. In the QUBO case, the computational complexity is $O(N^2)$.

If we have $M$ degenerate solutions, to obtain all of them, we will only have to apply the process $M$ times, so that we have a complexity of $O(M N^2 D^2)$.

However, we can improve execution time by reusing intermediate computations. To do so, when calculating the result of the first variable, we store the intermediate tensors $L^n$ that we see in Fig. \ref{fig:Contraction} b and c. In this way, since we only need to change the $n$-th node for a new $V^n$ tensor, as we explained for the iterative method, the rest of the tensor network, stored in the tensor $L^{n+1}$ will be the same. Therefore, we will only need to multiply the new $V^n$ tensor by the tensor $L^{n+1}$ that we had stored from the first iteration.

Since each of these multiplications has a cost $O(D)$, because the matrix is diagonal, and we have to do $O(N)$, the total cost of determining the variables from $1$ to $N-2$ will be $O(ND)$. Since the cost to determine the first is $O(N D^2)$, we will have a total computational complexity of $O(ND^2)$. The storage of the $L$-tensors in the first contraction requires space complexity $O(ND)$, so the total space complexity is $O(ND+D^2)$. Again, for Tensor QUDO the space complexity is $O(ND^2)$. In the QUBO case, the computational and space complexity is $O(N)$. If we have $M$ degenerate solutions, to obtain all of them, we will only have to apply the process $M$ times, so that we have a complexity of $O(M N D^2)$. Thus, it is determined that the algorithm has a linear cost in the number of variables and a quadratic cost in the variable dimensionality.

The algorithm can be parallelized, even without an improvement in computational complexity. For simplicity, we assume that we have enough units for computation that can compute in parallel, so we will analyze the parallel runtime with infinite processors. From this point on, we call computational complexity the main component of this parallel runtime. Our intention is to determine the vector after the Half Partial Trace without knowing the previous variables results. That is, we will compute these vectors in the first iteration of the tensor network contraction for every possible value of the previous variable and, after determining it, choose the corresponding vector. We will understand it better with the explicit algorithm. We define \textit{BigUnit} as a set of units that can compute in parallel the vector-matrix products, and \textit{SmallUnit} as the set of units that compose a BigUnit that computes in parallel each element of the result of the vector matrix product. The algorithm is as follows:
\begin{enumerate}
    \item We create each $Z$ tensor. Every tensor is computed in a different BigUnit, allowing us to compute them in parallel, and each element is computed in a different SmallUnit in its corresponding BigUnit, allowing their parallel computation. These are all the tensors required in the definition, and can be computed in a $O(1)$ time, but the last one that requires the summation of $D$ elements, taking $O(\log_2(D))$ time making pair sums. The computational complexity is $O(\log_2(D))$ and the space complexity is $O(ND^2)$. However, space complexity can be reduced to $O(N+D^2)$ if, instead of generating all $Z$-tensors in the first step, we only generate them just before contracting them.
    \item At the same time, in $D+1$ different BigUnits, we compute:
    \begin{itemize}
        \item We contract $Z^{N-1}$ with $Z^{N-2}$ to get $L^{N-2}$ in a BigUnit, computing each of its elements with a SmallUnit. Each element is computed in parallel and requires sum $D$ elements, which means it has computational complexity $O(\log_2(D))$ with pairs summation and space complexity $O(D^2)$.
        \item The contraction of $Z^{N-1}$ with $V^{N-2}$ for every possible value of $x_{N-3}$ to obtain all the vectors $P^{x_{N-2},x_{N-3}}$. Every contraction has computational complexity $O(1)$, because $V^{N-2}$ is a diagonal matrix, and they are performed in parallel in different BigUnits. Its space complexity is $O(D^2)$. 
        \item Now, for each value of $x_{N-3}$, we save the position of the largest component, obtaining the vector $X_{N-2}$. This vector component $X_{N-2,i}$ has the correct values for $x_{N-2}$ if the correct value of $x_{N-3}$ is $i$. Then,
        \begin{equation}
            X_{N-2,i} = \arg\max_j \left\lbrace P^{x_{N-2},i}_j\right\rbrace.
        \end{equation}
        This argmax computation has complexity $O(\log_2D)$ with pair comparisons, so it requires the same time as the contraction of $Z^{N-1}-Z^{N-2}$, making them parallel.
    \end{itemize}
    \item For $n$ from $N-2$ to $1$, defining $x_{-1}=0$:
    \begin{itemize}
        \item We contract $L^{n}$ with $Z^{n-1}$ to get $L^{n-1}$ in a BigUnit, computing each of its elements with a SmallUnit.
        \item The contraction of $L^{n}$ with $V^{n-1}$ for every possible value of $x_{n-2}$ to obtain all the vectors $P^{x_{n-1},x_{n-2}}$. 
        \item Now, for each value of $x_{n-2}$, we save the position of the largest component, obtaining the vector $X_{n-1}$. This vector component $X_{n-1,i}$ has the correct values for $x_{n-1}$ if the correct value of $x_{n-2}$ is $i$.
        \item We must perform these steps iteratively, so each iteration has a computational complexity of $O(\log_2D)$ and a space complexity of $O(D^2)$. Repetition of $N$ times increases computational complexity to $O(N\log_2D)$. The space complexity remains $O(ND+D^2)$, because we can overwrite the vectors $P$ at each step, and we only need to store the vectors $X$ of $D$ components.
    \end{itemize}
    
    \item We take $X_{0}$, which has only one component, because there is no previous variable, so $x_0 = X_{0,0}$. This step has computational complexity $O(1)$.
    \item For each $n$ from $1$ to $N-2$, we select $x_n = X_{n,x_{n-1}}$. Each iteration has computational complexity $O(1)$, so the total has complexity $O(N)$.
    \item The last variable value is determined by brute force, which requires $O(\log_2D)$ calculations.
\end{enumerate}

We call \textit{MeLoCoToN backtracking} to this last precomputation technique for the final backtracking evaluation of the solution. Taking into account all computational complexities and space complexities, the total computational complexity of the parallel algorithm is $O(N\log_2D)$ and the space complexity is $O(ND+D^2)$ (in Tensor QUDO it is $O(ND^2)$). So, the runtime of this algorithm scales linearly with the number of variables and logarithmically with their dimension. Tab.~\ref{tab:complex} shows the different computational and space complexity for both algorithms.

\begin{table*}[]
\begin{tabular}{|l|l|l|}
\hline
Algorithm                        & Computational Complexity & Space Complexity (QUDO$|$T-QUDO) \\
\hline
Brute Force                      & $O(D^N)$                   & $O(N)|O(ND^2)$             \\
\hline
LCQPTNS (this work)              & $O(N^2D^2)$                  & $O(N+D^2)|O(ND^2)$          \\
\hline
LCQPTNS with reuse (this work) & $O(ND^2)$                  & $O(ND+D^2)|O(ND^2)$          \\
\hline
Parallelized LCQPTNS (this work) & $O(N\log_2D)$                    & $O(ND+D^2)|O(ND^2)$       \\
\hline
\end{tabular}
\caption{Comparison of the different QUBO, QUDO and Tensor QUDO solvers for lineal chain with one-neighbor interaction for $N$ variables of values in the set $\lbrace 0, D-1\rbrace$.}
\label{tab:complex}
\end{table*}

\section{Experiments}\label{sec:experiments}
In this section, we introduce several experiments to analyze the performance of the proposed algorithms and compare it with the Google OR-TOOLS~\cite{ortools} and dimod~\cite{dimod} solvers. Tensor network algorithms are implemented in Python without parallelization, and all code is available in the GitHub repository. Every runtime experiment is performed three times and averaged to avoid fluctuations. In our implementation in NumPy, the value of $\tau$ for the new node increases to $\tau_n = \tau N / (N-n)$ to compensate for the reduction in the number of nodes. This considerably improves the quality of the results due to the finite precision of the numbers. We also normalize each tensor by its norm, and the intermediate computation tensors are also normalized. Instances are generated by uniform random numbers in the range $(-1,1)$. We also normalize the $w$ matrices/tensors and the $d$ vectors by their norms. We also normalize the $Z$ tensors and, for the Decimal implementation, if an overflow happens, we rescale $\tau$ until it does not diverge. However, the overflow problem remains, so we also implement the algorithm with the Decimal library for the QUDO algorithm, performing the multiplications with lists. This library allows to increase the precision of the numbers stored.  All experiments are performed in CPU, with an Intel(R) Core(TM) i7-14700HX 2.10 GHz and 16 GB RAM.

\begin{figure}
    \centering
    \includegraphics[width=1\linewidth]{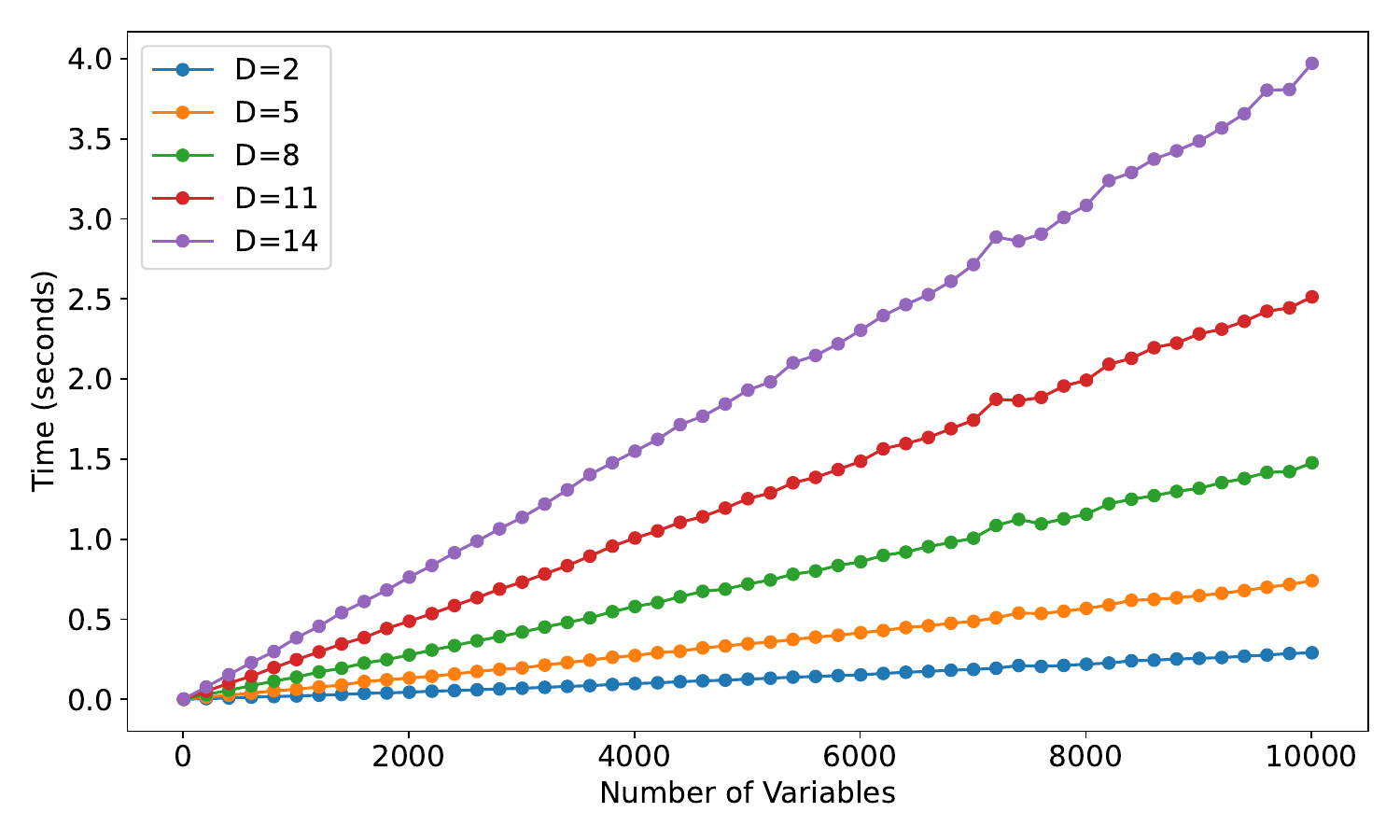}
    \caption{Run time for QUDO solver algorithm in NumPy against number of variables of the problem.}
    \label{fig:time_vs_variables_qudo}
\end{figure}

\begin{figure}
    \centering
    \includegraphics[width=\linewidth]{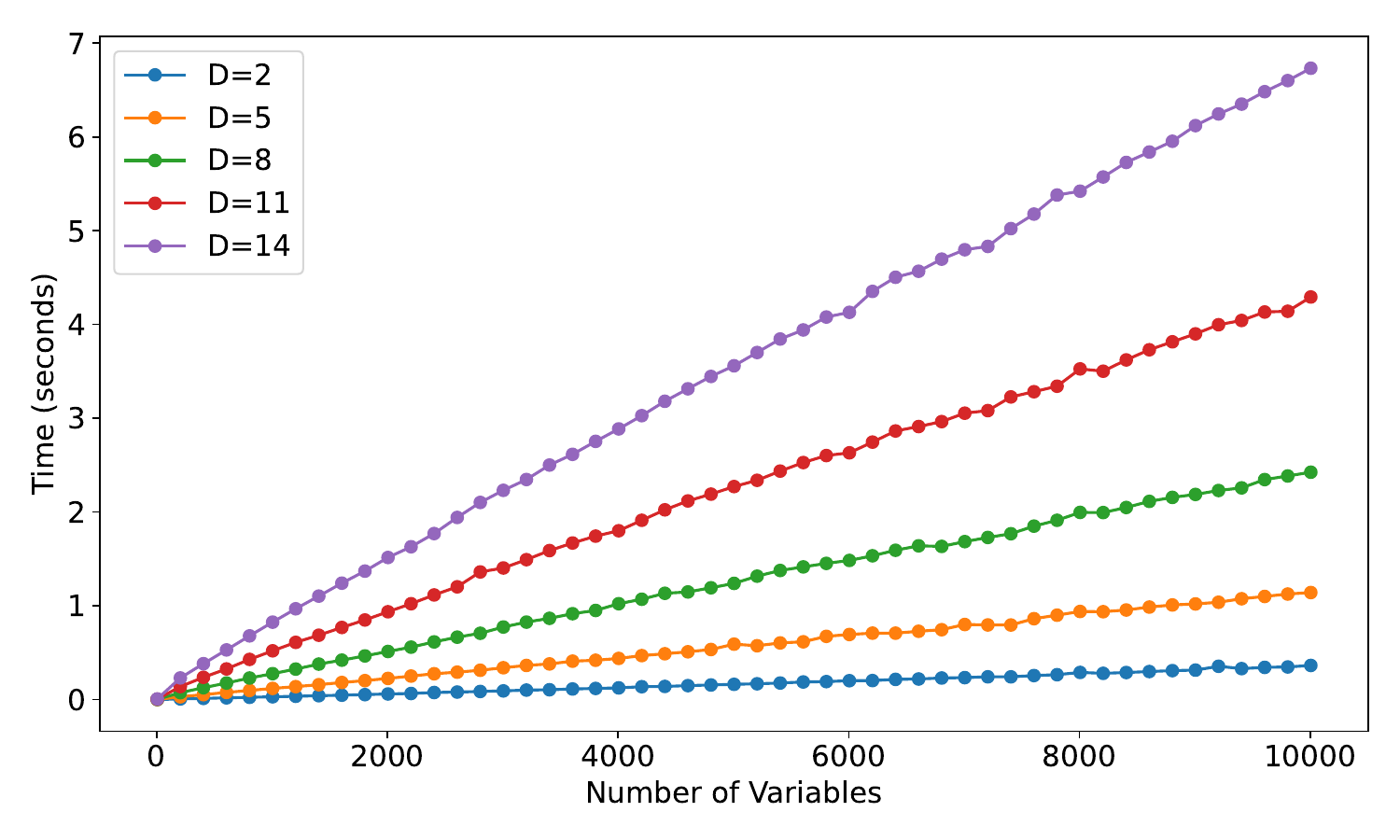}
    \caption{Run time for QUDO solver algorithm in Decimal against number of variables of the problem}
    \label{fig:time_vs_variables_qudo_decimal}
\end{figure}
\begin{figure}
    \centering
    \includegraphics[width=1\linewidth]{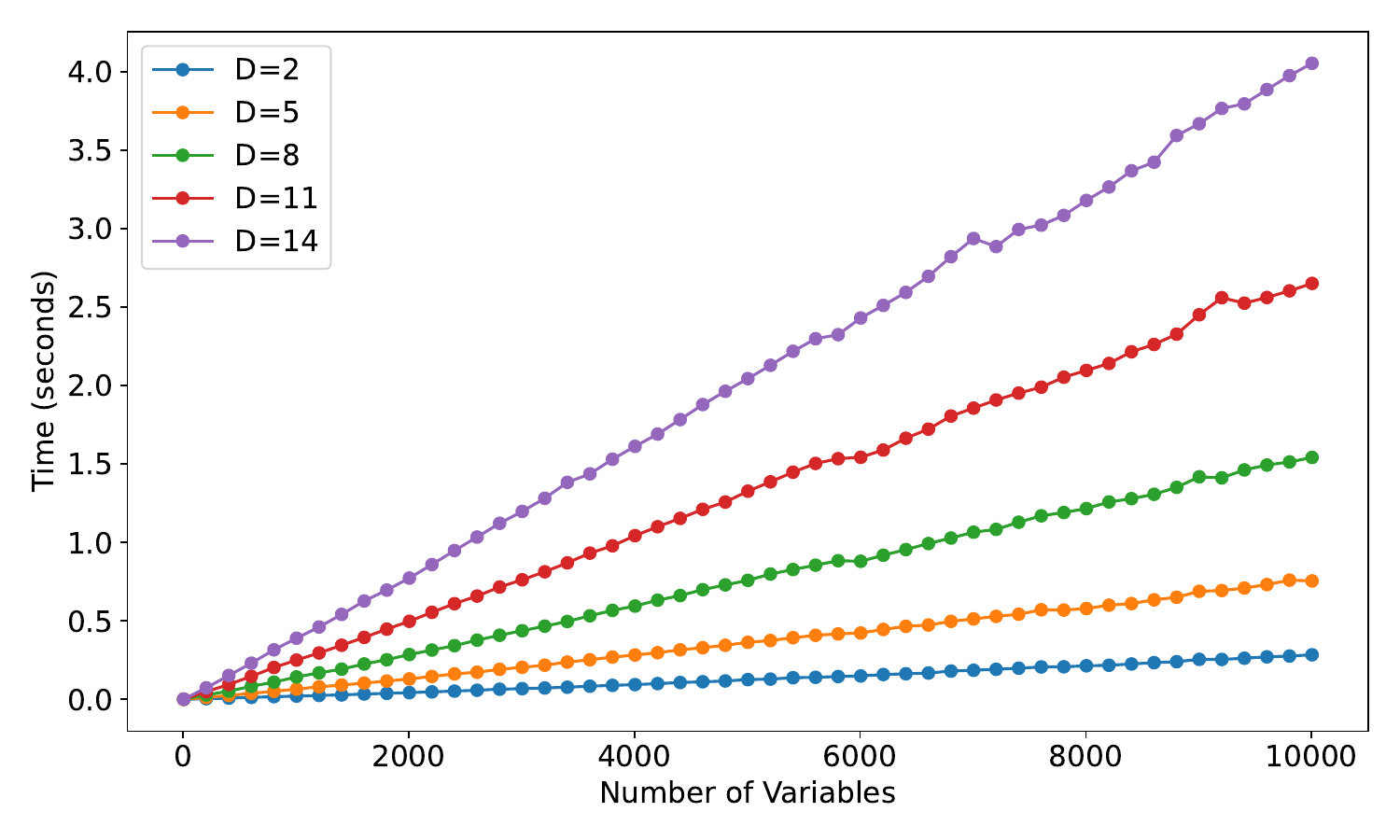}
    \caption{Run time for T-QUDO solver algorithm in NumPy against number of variables of the problem.}
    \label{fig:time_vs_variables_tensor_qudo}
\end{figure}

First, we test the scaling of the runtime with the number $N$ of variables. Figs.~\ref{fig:time_vs_variables_qudo}, \ref{fig:time_vs_variables_qudo_decimal} and \ref{fig:time_vs_variables_tensor_qudo} show the runtime for QUDO algorithms in Numpy and in Decimal, and for the Tensor QUDO algorithm, with the different number of variables. Both figures show that the runtime of all algorithms scales linearly with the number of variables, as we described. The Decimal implementation has a higher runtime, as expected, due to the higher precision and more bits operations.

\begin{figure}
    \centering
    \includegraphics[width=1\linewidth]{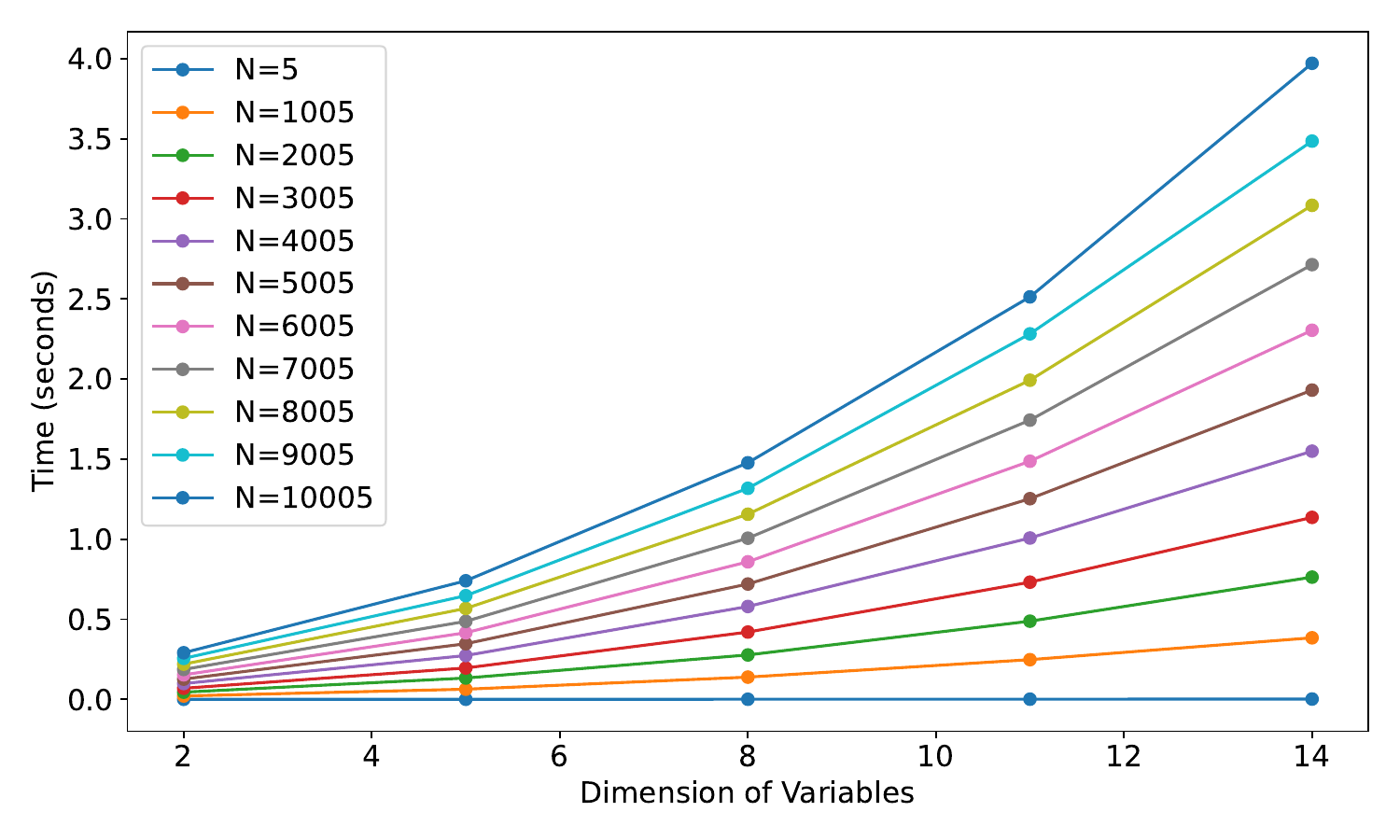}
    \caption{Run time for QUDO solver algorithm in NumPy against the dimension of variables of the problem.}
    \label{fig:time_vs_dimension_qudo}
\end{figure}

\begin{figure}
    \centering
    \includegraphics[width=1\linewidth]{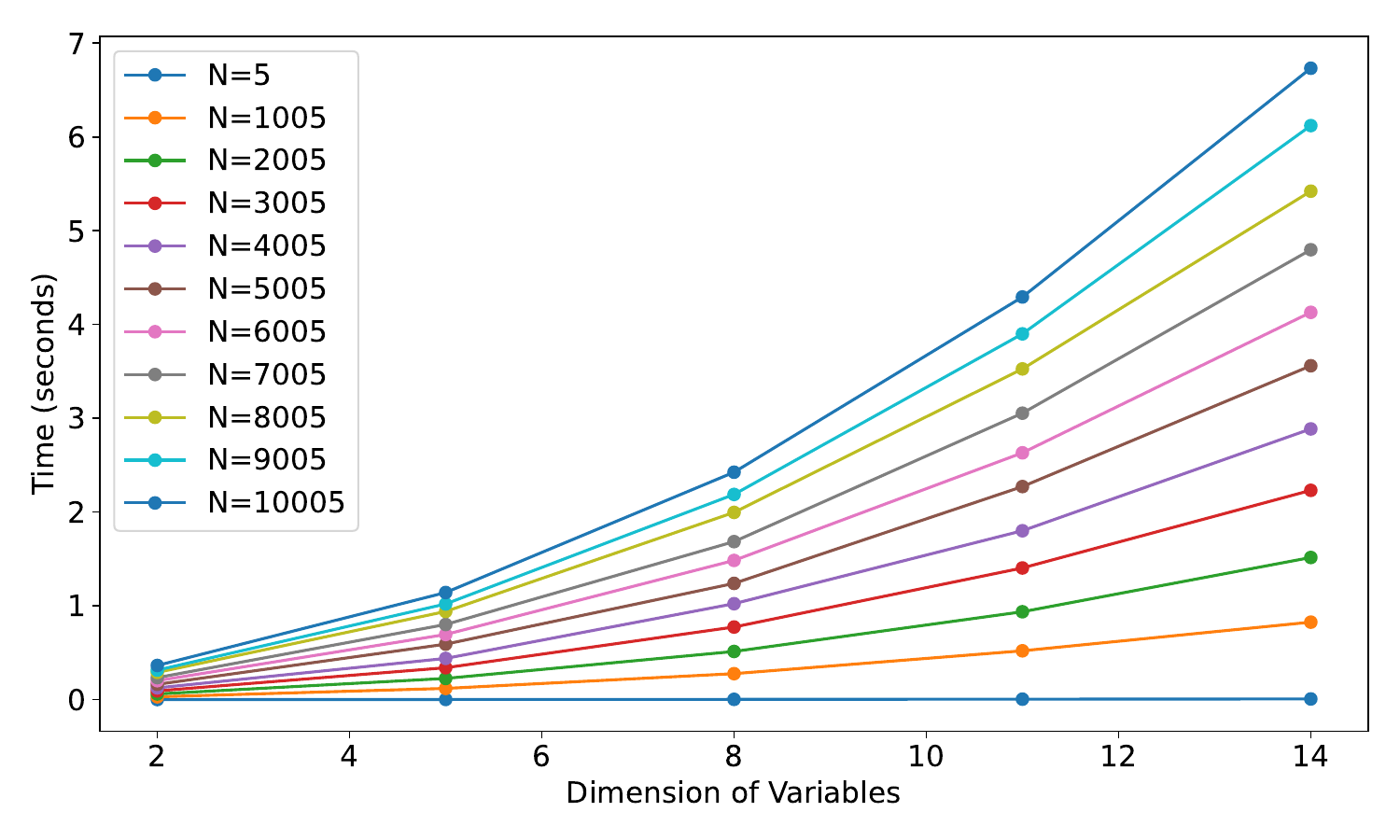}
    \caption{Run time for QUDO solver algorithm in Decimal against the dimension of variables of the problem.}
    \label{fig:time_vs_dimension_qudo_decimal}
\end{figure}

\begin{figure}
    \centering
    \includegraphics[width=1\linewidth]{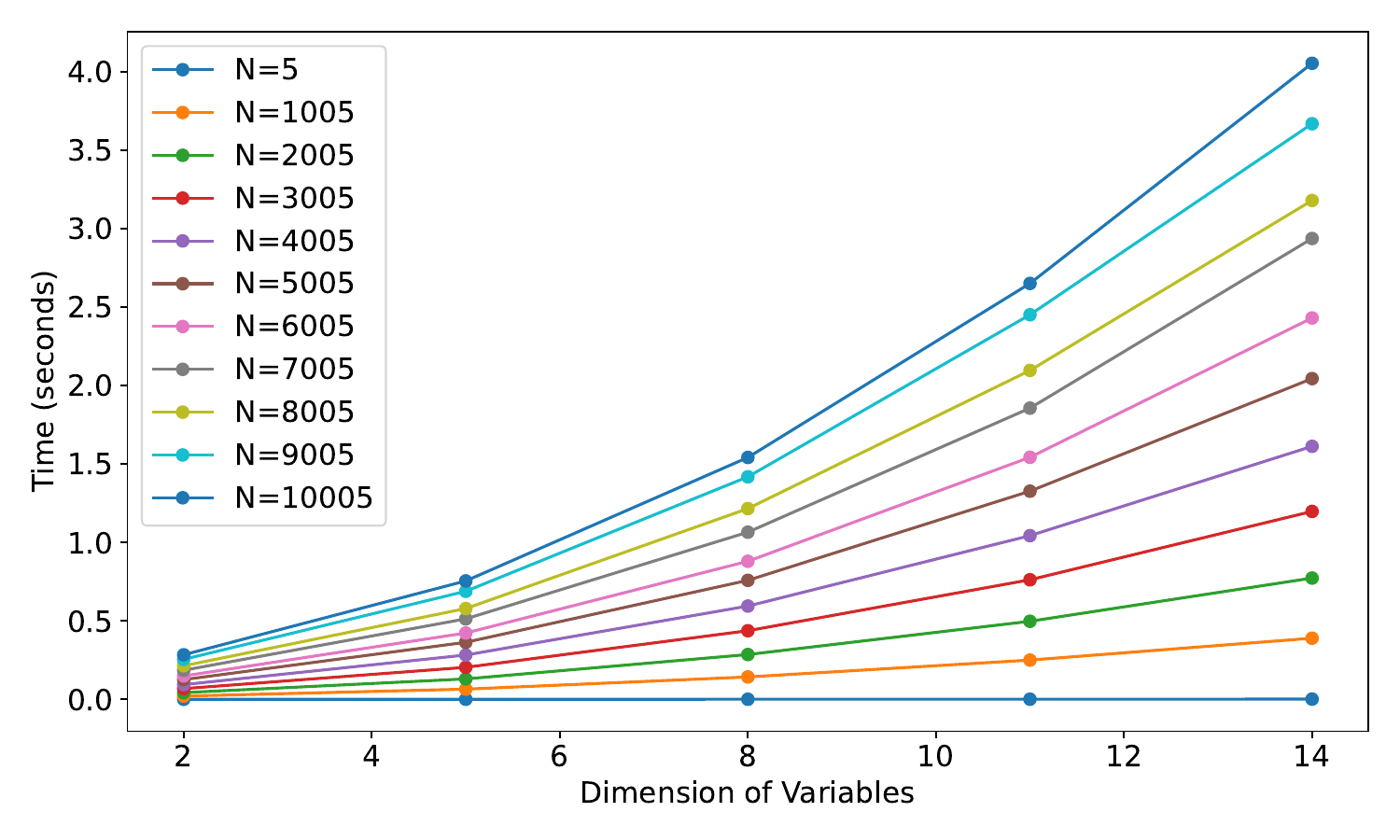}
    \caption{Run time for T-QUDO solver algorithm in NumPy against the dimension of variables of the problem.}
    \label{fig:time_vs_dimension_tensor_qudo}
\end{figure}

Now, we test the scaling of the runtime with the dimension $D$ of the variables. Figs.~\ref{fig:time_vs_dimension_qudo}, \ref{fig:time_vs_dimension_qudo_decimal} and \ref{fig:time_vs_dimension_tensor_qudo} show the runtime for QUDO algorithms in Numpy and in Decimal, and for the Tensor QUDO algorithm, with the different dimension of the variables. Both figures show that the runtime of all algorithms scales quadratically with the dimension of the variables, as we described. As before, the Decimal implementation has a higher runtime.

\begin{figure}
    \centering
    \includegraphics[width=1\linewidth]{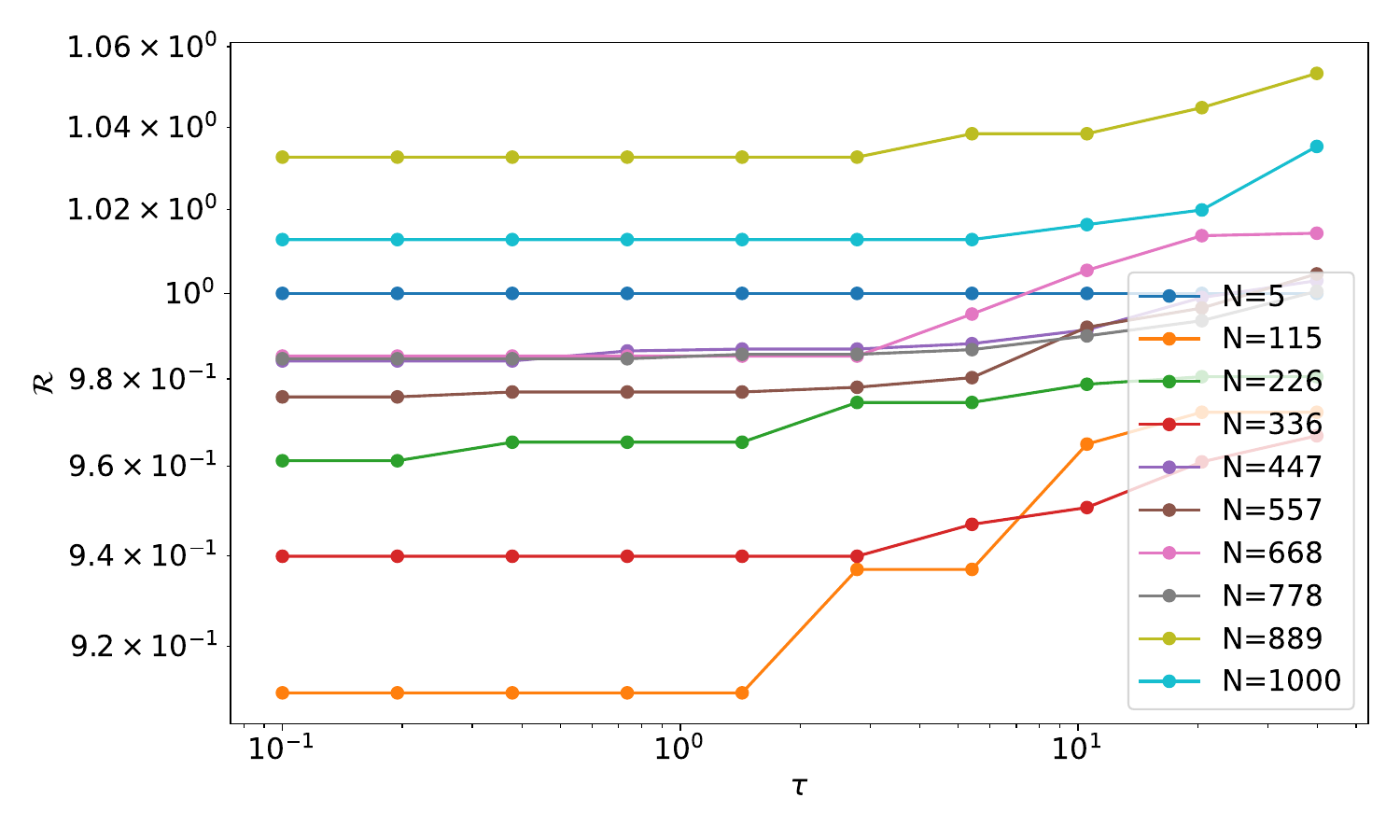}
    \caption{Ratio between the cost of the solutions of QUBO tensor networks solver in Numpy and QUBO simulated annealing solver or ORTOOLS solver against $\tau$.}
    \label{fig:ratio_vs_tau_qudo}
\end{figure}

\begin{figure}
    \centering
    \includegraphics[width=1\linewidth]{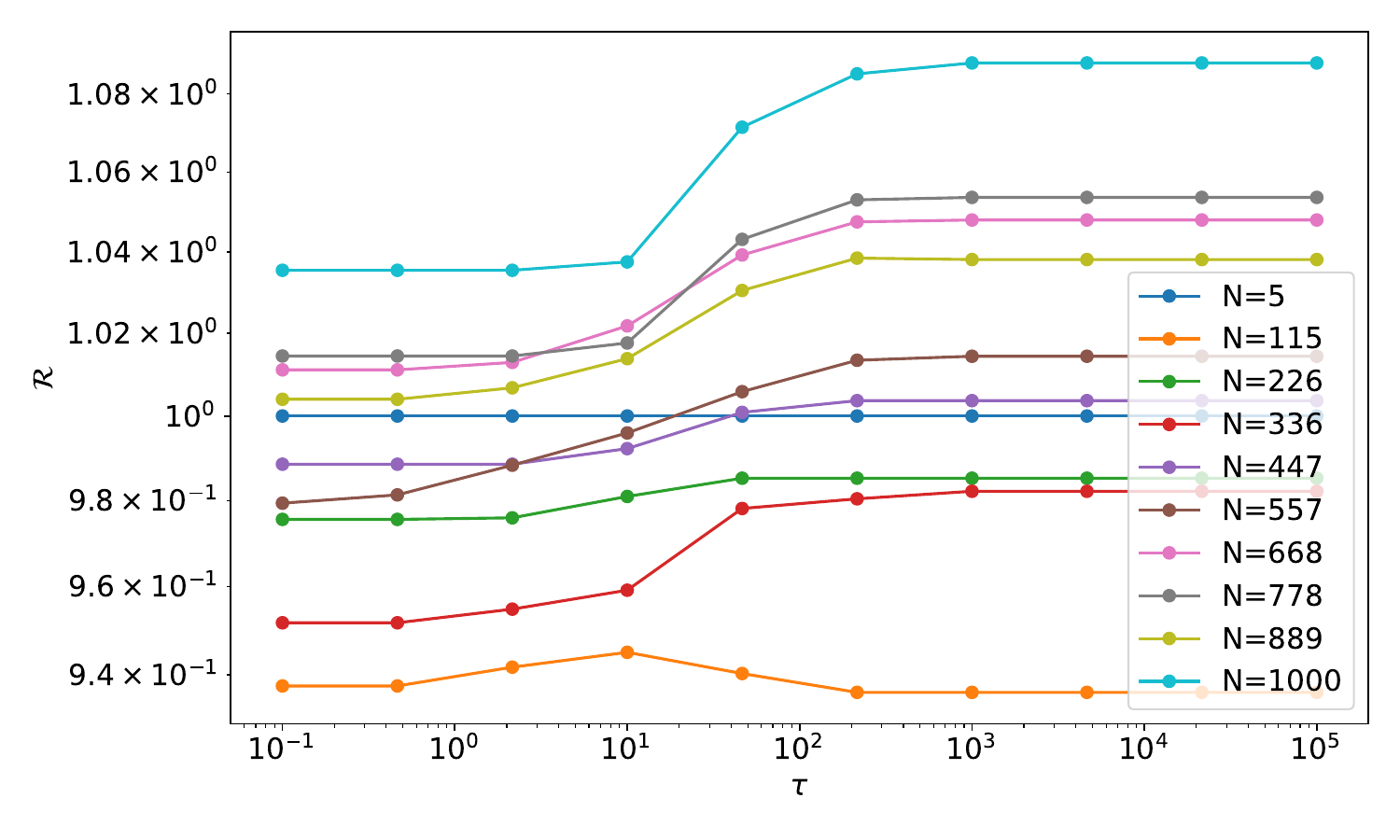}
    \caption{Ratio between the cost of the solutions of QUBO tensor networks solver in Decimal and QUBO simulated annealing solver or ORTOOLS solver against $\tau$.}
    \label{fig:ratio_vs_tau_qudo_decimal}
\end{figure}

Now, we compare the quality of the solutions with the value of $\tau$ for QUBO problems. We take as the correct solution the best provided by the Google OR-TOOLS and dimod solvers. The cost ratio is defined as
\begin{equation}
    \mathcal{R}=\frac{C(\vec{x}_{TN})}{C(\vec{x}_{opt})},
\end{equation}
being $C(\vec{x}_{TN})$ our tensor network solution. We perform the comparison for $D=2$ instances because it allows for a direct optimization by dimod's simulated annealing solver. For fair comparison, dimod and ORTOOLS are limited to a similar runtime to the tensor networks algorithms. Figs.~\ref{fig:ratio_vs_tau_qudo}, \ref{fig:ratio_vs_tau_qudo_decimal} show the cost ratio for QUDO algorithms in Numpy and in Decimal with the different values of $\tau$. We can see that, in general, there is an improvement in the ratio with the value of $\tau$, obtaining even better performance than dimod and ORTOOLS. We can see that, as we increase the number of variables, the tensor network results are better compared to the classical solvers. This shows the potential of this algorithm for larger instances.

\section{Conclusions}\label{sec:conclusions}
We have developed several algorithms in tensor networks to solve these lineal chain one-neighbor interaction quadratic problems efficiently, and provide the exact explicit equation that solves them. Moreover, we have implemented and tested them with other well-known classical algorithms and observed an improvement in the quality of the results. However, open questions remain. The first question is the minimal finite value of $\tau$ that results in the correct optimal solution. This depends on the dimensions and properties of the problem to solve and could be a future line of research. For example, a useful technique could be the iterative normalization described in \cite{Initi}. Another possible line will be the efficient resolution of more general QUBO, QUDO and Tensor QUDO problems with this methodology and its application to various industrial problems. In addition, it is possible to explore how to improve the computational complexity of the algorithm by inserting techniques such as quantization of the possible values of the tensor elements or taking advantage of the sparsity of the tensors. Finally, it is also interesting to test efficient implementations of the parallelized version of the algorithm we provided.

\section*{Acknowledgments}
The research leading to this paper has received funding from the Q4Real project (Quantum Computing for Real Industries), HAZITEK 2022, no. ZE-2022/00033.

\nocite{*}

\bibliography{apssamp}

\providecommand{\noopsort}[1]{}\providecommand{\singleletter}[1]{#1}%
\begin{thebibliography}{31}%
\makeatletter
\providecommand \@ifxundefined [1]{%
 \@ifx{#1\undefined}
}%
\providecommand \@ifnum [1]{%
 \ifnum #1\expandafter \@firstoftwo
 \else \expandafter \@secondoftwo
 \fi
}%
\providecommand \@ifx [1]{%
 \ifx #1\expandafter \@firstoftwo
 \else \expandafter \@secondoftwo
 \fi
}%
\providecommand \natexlab [1]{#1}%
\providecommand \enquote  [1]{``#1''}%
\providecommand \bibnamefont  [1]{#1}%
\providecommand \bibfnamefont [1]{#1}%
\providecommand \citenamefont [1]{#1}%
\providecommand \href@noop [0]{\@secondoftwo}%
\providecommand \href [0]{\begingroup \@sanitize@url \@href}%
\providecommand \@href[1]{\@@startlink{#1}\@@href}%
\providecommand \@@href[1]{\endgroup#1\@@endlink}%
\providecommand \@sanitize@url [0]{\catcode `\\12\catcode `\$12\catcode `\&12\catcode `\#12\catcode `\^12\catcode `\_12\catcode `\%12\relax}%
\providecommand \@@startlink[1]{}%
\providecommand \@@endlink[0]{}%
\providecommand \url  [0]{\begingroup\@sanitize@url \@url }%
\providecommand \@url [1]{\endgroup\@href {#1}{\urlprefix }}%
\providecommand \urlprefix  [0]{URL }%
\providecommand \Eprint [0]{\href }%
\providecommand \doibase [0]{https://doi.org/}%
\providecommand \selectlanguage [0]{\@gobble}%
\providecommand \bibinfo  [0]{\@secondoftwo}%
\providecommand \bibfield  [0]{\@secondoftwo}%
\providecommand \translation [1]{[#1]}%
\providecommand \BibitemOpen [0]{}%
\providecommand \bibitemStop [0]{}%
\providecommand \bibitemNoStop [0]{.\EOS\space}%
\providecommand \EOS [0]{\spacefactor3000\relax}%
\providecommand \BibitemShut  [1]{\csname bibitem#1\endcsname}%
\let\auto@bib@innerbib\@empty
\bibitem [{\citenamefont {Glover}\ \emph {et~al.}(2019)\citenamefont {Glover}, \citenamefont {Kochenberger},\ and\ \citenamefont {Du}}]{QUBO}%
  \BibitemOpen
  \bibfield  {author} {\bibinfo {author} {\bibfnamefont {F.}~\bibnamefont {Glover}}, \bibinfo {author} {\bibfnamefont {G.}~\bibnamefont {Kochenberger}},\ and\ \bibinfo {author} {\bibfnamefont {Y.}~\bibnamefont {Du}},\ }\href@noop {} {\bibinfo {title} {A tutorial on formulating and using qubo models}} (\bibinfo {year} {2019}),\ \Eprint {https://arxiv.org/abs/1811.11538} {arXiv:1811.11538 [cs.DS]} \BibitemShut {NoStop}%
\bibitem [{\citenamefont {Sales}\ and\ \citenamefont {Araos}(2023)}]{Logist}%
  \BibitemOpen
  \bibfield  {author} {\bibinfo {author} {\bibfnamefont {J.~F.~A.}\ \bibnamefont {Sales}}\ and\ \bibinfo {author} {\bibfnamefont {R.~A.~P.}\ \bibnamefont {Araos}},\ }\href@noop {} {\bibinfo {title} {Adiabatic quantum computing for logistic transport optimization}} (\bibinfo {year} {2023}),\ \Eprint {https://arxiv.org/abs/2301.07691} {arXiv:2301.07691 [quant-ph]} \BibitemShut {NoStop}%
\bibitem [{\citenamefont {Romero}\ \emph {et~al.}(2023)\citenamefont {Romero}, \citenamefont {Osaba}, \citenamefont {Villar-Rodriguez},\ and\ \citenamefont {Asla}}]{Logist2}%
  \BibitemOpen
  \bibfield  {author} {\bibinfo {author} {\bibfnamefont {S.~V.}\ \bibnamefont {Romero}}, \bibinfo {author} {\bibfnamefont {E.}~\bibnamefont {Osaba}}, \bibinfo {author} {\bibfnamefont {E.}~\bibnamefont {Villar-Rodriguez}},\ and\ \bibinfo {author} {\bibfnamefont {A.}~\bibnamefont {Asla}},\ }\bibfield  {title} {\bibinfo {title} {Solving logistic-oriented bin packing problems through a hybrid quantum-classical approach},\ }in\ \href {https://doi.org/10.1109/ITSC57777.2023.10422581} {\emph {\bibinfo {booktitle} {2023 IEEE 26th International Conference on Intelligent Transportation Systems (ITSC)}}}\ (\bibinfo {year} {2023})\ pp.\ \bibinfo {pages} {2239--2245}\BibitemShut {NoStop}%
\bibitem [{\citenamefont {Matsumori}\ \emph {et~al.}(2022)\citenamefont {Matsumori}, \citenamefont {Taki},\ and\ \citenamefont {Kadowaki}}]{Engin}%
  \BibitemOpen
  \bibfield  {author} {\bibinfo {author} {\bibfnamefont {T.}~\bibnamefont {Matsumori}}, \bibinfo {author} {\bibfnamefont {M.}~\bibnamefont {Taki}},\ and\ \bibinfo {author} {\bibfnamefont {T.}~\bibnamefont {Kadowaki}},\ }\bibfield  {title} {\bibinfo {title} {Application of qubo solver using black-box optimization to structural design for resonance avoidance},\ }\href@noop {} {\bibfield  {journal} {\bibinfo  {journal} {Scientific Reports}\ }\textbf {\bibinfo {volume} {12}},\ \bibinfo {pages} {12143} (\bibinfo {year} {2022})}\BibitemShut {NoStop}%
\bibitem [{\citenamefont {Zaborniak}\ \emph {et~al.}(2022)\citenamefont {Zaborniak}, \citenamefont {Giraldo}, \citenamefont {Müller}, \citenamefont {Jabbari},\ and\ \citenamefont {Stege}}]{Biol}%
  \BibitemOpen
  \bibfield  {author} {\bibinfo {author} {\bibfnamefont {T.}~\bibnamefont {Zaborniak}}, \bibinfo {author} {\bibfnamefont {J.}~\bibnamefont {Giraldo}}, \bibinfo {author} {\bibfnamefont {H.}~\bibnamefont {Müller}}, \bibinfo {author} {\bibfnamefont {H.}~\bibnamefont {Jabbari}},\ and\ \bibinfo {author} {\bibfnamefont {U.}~\bibnamefont {Stege}},\ }\bibfield  {title} {\bibinfo {title} {A qubo model of the rna folding problem optimized by variational hybrid quantum annealing},\ }in\ \href {https://doi.org/10.1109/QCE53715.2022.00037} {\emph {\bibinfo {booktitle} {2022 IEEE International Conference on Quantum Computing and Engineering (QCE)}}}\ (\bibinfo {year} {2022})\ pp.\ \bibinfo {pages} {174--185}\BibitemShut {NoStop}%
\bibitem [{\citenamefont {Mattesi}\ \emph {et~al.}(2024)\citenamefont {Mattesi}, \citenamefont {Asproni}, \citenamefont {Mattia}, \citenamefont {Tufano}, \citenamefont {Ranieri}, \citenamefont {Caputo},\ and\ \citenamefont {Corbelletto}}]{Finances}%
  \BibitemOpen
  \bibfield  {author} {\bibinfo {author} {\bibfnamefont {M.}~\bibnamefont {Mattesi}}, \bibinfo {author} {\bibfnamefont {L.}~\bibnamefont {Asproni}}, \bibinfo {author} {\bibfnamefont {C.}~\bibnamefont {Mattia}}, \bibinfo {author} {\bibfnamefont {S.}~\bibnamefont {Tufano}}, \bibinfo {author} {\bibfnamefont {G.}~\bibnamefont {Ranieri}}, \bibinfo {author} {\bibfnamefont {D.}~\bibnamefont {Caputo}},\ and\ \bibinfo {author} {\bibfnamefont {D.}~\bibnamefont {Corbelletto}},\ }\href@noop {} {\bibinfo {title} {Diversifying investments and maximizing sharpe ratio: a novel qubo formulation}} (\bibinfo {year} {2024}),\ \Eprint {https://arxiv.org/abs/2302.12291} {arXiv:2302.12291 [quant-ph]} \BibitemShut {NoStop}%
\bibitem [{\citenamefont {Mata~Ali}()}]{t_qudo}%
  \BibitemOpen
  \bibfield  {author} {\bibinfo {author} {\bibfnamefont {A.}~\bibnamefont {Mata~Ali}},\ }\href@noop {} {\bibinfo {title} {Introduction to qudo, tensor qudo and hobo formulations: Qudits, equivalences, knapsack problem, traveling salesman problem and combinatorial games}}\BibitemShut {NoStop}%
\bibitem [{\citenamefont {Yasuoka}(2022)}]{Complex}%
  \BibitemOpen
  \bibfield  {author} {\bibinfo {author} {\bibfnamefont {H.}~\bibnamefont {Yasuoka}},\ }\href@noop {} {\bibinfo {title} {Computational complexity of quadratic unconstrained binary optimization}} (\bibinfo {year} {2022}),\ \Eprint {https://arxiv.org/abs/2109.10048} {arXiv:2109.10048 [cs.CC]} \BibitemShut {NoStop}%
\bibitem [{\citenamefont {{\c{C}}ela}\ and\ \citenamefont {Punnen}(2022)}]{Polynomial}%
  \BibitemOpen
  \bibfield  {author} {\bibinfo {author} {\bibfnamefont {E.}~\bibnamefont {{\c{C}}ela}}\ and\ \bibinfo {author} {\bibfnamefont {A.~P.}\ \bibnamefont {Punnen}},\ }\bibinfo {title} {Complexity and polynomially solvable special cases of qubo},\ in\ \href {https://doi.org/10.1007/978-3-031-04520-2_3} {\emph {\bibinfo {booktitle} {The Quadratic Unconstrained Binary Optimization Problem: Theory, Algorithms, and Applications}}},\ \bibinfo {editor} {edited by\ \bibinfo {editor} {\bibfnamefont {A.~P.}\ \bibnamefont {Punnen}}}\ (\bibinfo  {publisher} {Springer International Publishing},\ \bibinfo {address} {Cham},\ \bibinfo {year} {2022})\ pp.\ \bibinfo {pages} {57--95}\BibitemShut {NoStop}%
\bibitem [{\citenamefont {Allemand}\ \emph {et~al.}(2001)\citenamefont {Allemand}, \citenamefont {Fukuda}, \citenamefont {Liebling},\ and\ \citenamefont {Steiner}}]{Zero-One}%
  \BibitemOpen
  \bibfield  {author} {\bibinfo {author} {\bibfnamefont {K.}~\bibnamefont {Allemand}}, \bibinfo {author} {\bibfnamefont {K.}~\bibnamefont {Fukuda}}, \bibinfo {author} {\bibfnamefont {T.~M.}\ \bibnamefont {Liebling}},\ and\ \bibinfo {author} {\bibfnamefont {E.}~\bibnamefont {Steiner}},\ }\bibfield  {title} {\bibinfo {title} {A polynomial case of unconstrained zero-one quadratic optimization},\ }\href {https://doi.org/10.1007/s101070100233} {\bibfield  {journal} {\bibinfo  {journal} {Mathematical Programming}\ }\textbf {\bibinfo {volume} {91}},\ \bibinfo {pages} {49} (\bibinfo {year} {2001})}\BibitemShut {NoStop}%
\bibitem [{\citenamefont {Nakano}\ \emph {et~al.}(2023)\citenamefont {Nakano}, \citenamefont {Takafuji}, \citenamefont {Ito}, \citenamefont {Yazane}, \citenamefont {Yano}, \citenamefont {Ozaki}, \citenamefont {Katsuki},\ and\ \citenamefont {Mori}}]{GPU}%
  \BibitemOpen
  \bibfield  {author} {\bibinfo {author} {\bibfnamefont {K.}~\bibnamefont {Nakano}}, \bibinfo {author} {\bibfnamefont {D.}~\bibnamefont {Takafuji}}, \bibinfo {author} {\bibfnamefont {Y.}~\bibnamefont {Ito}}, \bibinfo {author} {\bibfnamefont {T.}~\bibnamefont {Yazane}}, \bibinfo {author} {\bibfnamefont {J.}~\bibnamefont {Yano}}, \bibinfo {author} {\bibfnamefont {S.}~\bibnamefont {Ozaki}}, \bibinfo {author} {\bibfnamefont {R.}~\bibnamefont {Katsuki}},\ and\ \bibinfo {author} {\bibfnamefont {R.}~\bibnamefont {Mori}},\ }\bibfield  {title} {\bibinfo {title} {Diverse adaptive bulk search: a framework for solving qubo problems on multiple gpus},\ }in\ \href {https://doi.org/10.1109/IPDPSW59300.2023.00060} {\emph {\bibinfo {booktitle} {2023 IEEE International Parallel and Distributed Processing Symposium Workshops (IPDPSW)}}}\ (\bibinfo {year} {2023})\ pp.\ \bibinfo {pages} {314--325}\BibitemShut {NoStop}%
\bibitem [{\citenamefont {Aharonov}\ \emph {et~al.}(2010)\citenamefont {Aharonov}, \citenamefont {Arad},\ and\ \citenamefont {Irani}}]{1D_Ground}%
  \BibitemOpen
  \bibfield  {author} {\bibinfo {author} {\bibfnamefont {D.}~\bibnamefont {Aharonov}}, \bibinfo {author} {\bibfnamefont {I.}~\bibnamefont {Arad}},\ and\ \bibinfo {author} {\bibfnamefont {S.}~\bibnamefont {Irani}},\ }\bibfield  {title} {\bibinfo {title} {Efficient algorithm for approximating one-dimensional ground states},\ }\href {https://doi.org/10.1103/PhysRevA.82.012315} {\bibfield  {journal} {\bibinfo  {journal} {Phys. Rev. A}\ }\textbf {\bibinfo {volume} {82}},\ \bibinfo {pages} {012315} (\bibinfo {year} {2010})}\BibitemShut {NoStop}%
\bibitem [{\citenamefont {Aramon}\ \emph {et~al.}(2019)\citenamefont {Aramon}, \citenamefont {Rosenberg}, \citenamefont {Valiante}, \citenamefont {Miyazawa}, \citenamefont {Tamura},\ and\ \citenamefont {Katzgraber}}]{Digital}%
  \BibitemOpen
  \bibfield  {author} {\bibinfo {author} {\bibfnamefont {M.}~\bibnamefont {Aramon}}, \bibinfo {author} {\bibfnamefont {G.}~\bibnamefont {Rosenberg}}, \bibinfo {author} {\bibfnamefont {E.}~\bibnamefont {Valiante}}, \bibinfo {author} {\bibfnamefont {T.}~\bibnamefont {Miyazawa}}, \bibinfo {author} {\bibfnamefont {H.}~\bibnamefont {Tamura}},\ and\ \bibinfo {author} {\bibfnamefont {H.~G.}\ \bibnamefont {Katzgraber}},\ }\bibfield  {title} {\bibinfo {title} {Physics-inspired optimization for quadratic unconstrained problems using a digital annealer},\ }\bibfield  {journal} {\bibinfo  {journal} {Frontiers in Physics}\ }\textbf {\bibinfo {volume} {7}},\ \href {https://doi.org/10.3389/fphy.2019.00048} {10.3389/fphy.2019.00048} (\bibinfo {year} {2019})\BibitemShut {NoStop}%
\bibitem [{\citenamefont {Farhi}\ \emph {et~al.}(2014)\citenamefont {Farhi}, \citenamefont {Goldstone},\ and\ \citenamefont {Gutmann}}]{qaoa}%
  \BibitemOpen
  \bibfield  {author} {\bibinfo {author} {\bibfnamefont {E.}~\bibnamefont {Farhi}}, \bibinfo {author} {\bibfnamefont {J.}~\bibnamefont {Goldstone}},\ and\ \bibinfo {author} {\bibfnamefont {S.}~\bibnamefont {Gutmann}},\ }\href {https://arxiv.org/abs/1411.4028} {\bibinfo {title} {A quantum approximate optimization algorithm}} (\bibinfo {year} {2014}),\ \Eprint {https://arxiv.org/abs/1411.4028} {arXiv:1411.4028 [quant-ph]} \BibitemShut {NoStop}%
\bibitem [{\citenamefont {Ákos Nagy}\ \emph {et~al.}(2023)\citenamefont {Ákos Nagy}, \citenamefont {Park}, \citenamefont {Zhang}, \citenamefont {Acharya},\ and\ \citenamefont {Khan}}]{Grover}%
  \BibitemOpen
  \bibfield  {author} {\bibinfo {author} {\bibnamefont {Ákos Nagy}}, \bibinfo {author} {\bibfnamefont {J.}~\bibnamefont {Park}}, \bibinfo {author} {\bibfnamefont {C.}~\bibnamefont {Zhang}}, \bibinfo {author} {\bibfnamefont {A.}~\bibnamefont {Acharya}},\ and\ \bibinfo {author} {\bibfnamefont {A.}~\bibnamefont {Khan}},\ }\href@noop {} {\bibinfo {title} {Fixed-point grover adaptive search for qubo problems}} (\bibinfo {year} {2023}),\ \Eprint {https://arxiv.org/abs/2311.05592} {arXiv:2311.05592 [quant-ph]} \BibitemShut {NoStop}%
\bibitem [{\citenamefont {Delgado}\ \emph {et~al.}(2022)\citenamefont {Delgado}, \citenamefont {Markaida}, \citenamefont {De~Leceta},\ and\ \citenamefont {Uriarte}}]{Hobbit}%
  \BibitemOpen
  \bibfield  {author} {\bibinfo {author} {\bibfnamefont {I.~P.}\ \bibnamefont {Delgado}}, \bibinfo {author} {\bibfnamefont {B.~G.}\ \bibnamefont {Markaida}}, \bibinfo {author} {\bibfnamefont {A.~M.~F.}\ \bibnamefont {De~Leceta}},\ and\ \bibinfo {author} {\bibfnamefont {J.~A.~O.}\ \bibnamefont {Uriarte}},\ }\bibfield  {title} {\bibinfo {title} {Quantum hobbit routing: Annealer implementation of generalized travelling salesperson problem},\ }in\ \href {https://doi.org/10.1109/SSCI51031.2022.10022127} {\emph {\bibinfo {booktitle} {2022 IEEE Symposium Series on Computational Intelligence (SSCI)}}}\ (\bibinfo {year} {2022})\ pp.\ \bibinfo {pages} {923--929}\BibitemShut {NoStop}%
\bibitem [{\citenamefont {Pramanik}\ and\ \citenamefont {Chandra}(2021)}]{QUDO_QAOA}%
  \BibitemOpen
  \bibfield  {author} {\bibinfo {author} {\bibfnamefont {S.}~\bibnamefont {Pramanik}}\ and\ \bibinfo {author} {\bibfnamefont {M.~G.}\ \bibnamefont {Chandra}},\ }\href@noop {} {\bibinfo {title} {Quantum-assisted graph clustering and quadratic unconstrained d-ary optimisation}} (\bibinfo {year} {2021}),\ \Eprint {https://arxiv.org/abs/2004.02608} {arXiv:2004.02608 [quant-ph]} \BibitemShut {NoStop}%
\bibitem [{\citenamefont {Yan}\ \emph {et~al.}(2023)\citenamefont {Yan}, \citenamefont {Du}, \citenamefont {Chen},\ and\ \citenamefont {Ma}}]{nisq}%
  \BibitemOpen
  \bibfield  {author} {\bibinfo {author} {\bibfnamefont {Y.}~\bibnamefont {Yan}}, \bibinfo {author} {\bibfnamefont {Z.}~\bibnamefont {Du}}, \bibinfo {author} {\bibfnamefont {J.}~\bibnamefont {Chen}},\ and\ \bibinfo {author} {\bibfnamefont {X.}~\bibnamefont {Ma}},\ }\href {https://arxiv.org/abs/2306.02836} {\bibinfo {title} {Limitations of noisy quantum devices in computational and entangling power}} (\bibinfo {year} {2023}),\ \Eprint {https://arxiv.org/abs/2306.02836} {arXiv:2306.02836 [quant-ph]} \BibitemShut {NoStop}%
\bibitem [{\citenamefont {Biamonte}\ and\ \citenamefont {Bergholm}(2017)}]{Tensor_Network}%
  \BibitemOpen
  \bibfield  {author} {\bibinfo {author} {\bibfnamefont {J.}~\bibnamefont {Biamonte}}\ and\ \bibinfo {author} {\bibfnamefont {V.}~\bibnamefont {Bergholm}},\ }\href@noop {} {\bibinfo {title} {Tensor networks in a nutshell}} (\bibinfo {year} {2017}),\ \Eprint {https://arxiv.org/abs/1708.00006} {arXiv:1708.00006 [quant-ph]} \BibitemShut {NoStop}%
\bibitem [{\citenamefont {Orús}(2014)}]{orus_tn}%
  \BibitemOpen
  \bibfield  {author} {\bibinfo {author} {\bibfnamefont {R.}~\bibnamefont {Orús}},\ }\bibfield  {title} {\bibinfo {title} {A practical introduction to tensor networks: Matrix product states and projected entangled pair states},\ }\href {https://doi.org/10.1016/j.aop.2014.06.013} {\bibfield  {journal} {\bibinfo  {journal} {Annals of Physics}\ }\textbf {\bibinfo {volume} {349}},\ \bibinfo {pages} {117–158} (\bibinfo {year} {2014})}\BibitemShut {NoStop}%
\bibitem [{\citenamefont {Novikov}\ \emph {et~al.}(2015)\citenamefont {Novikov}, \citenamefont {Podoprikhin}, \citenamefont {Osokin},\ and\ \citenamefont {Vetrov}}]{TNN}%
  \BibitemOpen
  \bibfield  {author} {\bibinfo {author} {\bibfnamefont {A.}~\bibnamefont {Novikov}}, \bibinfo {author} {\bibfnamefont {D.}~\bibnamefont {Podoprikhin}}, \bibinfo {author} {\bibfnamefont {A.}~\bibnamefont {Osokin}},\ and\ \bibinfo {author} {\bibfnamefont {D.}~\bibnamefont {Vetrov}},\ }\href {https://arxiv.org/abs/1509.06569} {\bibinfo {title} {Tensorizing neural networks}} (\bibinfo {year} {2015}),\ \Eprint {https://arxiv.org/abs/1509.06569} {arXiv:1509.06569 [cs.LG]} \BibitemShut {NoStop}%
\bibitem [{\citenamefont {Tomut}\ \emph {et~al.}(2024)\citenamefont {Tomut}, \citenamefont {Jahromi}, \citenamefont {Sarkar}, \citenamefont {Kurt}, \citenamefont {Singh}, \citenamefont {Ishtiaq}, \citenamefont {Muñoz}, \citenamefont {Bajaj}, \citenamefont {Elborady}, \citenamefont {del Bimbo}, \citenamefont {Alizadeh}, \citenamefont {Montero}, \citenamefont {Martin-Ramiro}, \citenamefont {Ibrahim}, \citenamefont {Alaoui}, \citenamefont {Malcolm}, \citenamefont {Mugel},\ and\ \citenamefont {Orus}}]{compactifai}%
  \BibitemOpen
  \bibfield  {author} {\bibinfo {author} {\bibfnamefont {A.}~\bibnamefont {Tomut}}, \bibinfo {author} {\bibfnamefont {S.~S.}\ \bibnamefont {Jahromi}}, \bibinfo {author} {\bibfnamefont {A.}~\bibnamefont {Sarkar}}, \bibinfo {author} {\bibfnamefont {U.}~\bibnamefont {Kurt}}, \bibinfo {author} {\bibfnamefont {S.}~\bibnamefont {Singh}}, \bibinfo {author} {\bibfnamefont {F.}~\bibnamefont {Ishtiaq}}, \bibinfo {author} {\bibfnamefont {C.}~\bibnamefont {Muñoz}}, \bibinfo {author} {\bibfnamefont {P.~S.}\ \bibnamefont {Bajaj}}, \bibinfo {author} {\bibfnamefont {A.}~\bibnamefont {Elborady}}, \bibinfo {author} {\bibfnamefont {G.}~\bibnamefont {del Bimbo}}, \bibinfo {author} {\bibfnamefont {M.}~\bibnamefont {Alizadeh}}, \bibinfo {author} {\bibfnamefont {D.}~\bibnamefont {Montero}}, \bibinfo {author} {\bibfnamefont {P.}~\bibnamefont {Martin-Ramiro}}, \bibinfo {author} {\bibfnamefont {M.}~\bibnamefont {Ibrahim}}, \bibinfo {author} {\bibfnamefont {O.~T.}\ \bibnamefont {Alaoui}}, \bibinfo {author} {\bibfnamefont
  {J.}~\bibnamefont {Malcolm}}, \bibinfo {author} {\bibfnamefont {S.}~\bibnamefont {Mugel}},\ and\ \bibinfo {author} {\bibfnamefont {R.}~\bibnamefont {Orus}},\ }\href {https://arxiv.org/abs/2401.14109} {\bibinfo {title} {Compactifai: Extreme compression of large language models using quantum-inspired tensor networks}} (\bibinfo {year} {2024}),\ \Eprint {https://arxiv.org/abs/2401.14109} {arXiv:2401.14109 [cs.CL]} \BibitemShut {NoStop}%
\bibitem [{\citenamefont {Seitz}\ \emph {et~al.}(2023)\citenamefont {Seitz}, \citenamefont {Medina}, \citenamefont {Cruz}, \citenamefont {Huang},\ and\ \citenamefont {Mendl}}]{Simul_tn}%
  \BibitemOpen
  \bibfield  {author} {\bibinfo {author} {\bibfnamefont {P.}~\bibnamefont {Seitz}}, \bibinfo {author} {\bibfnamefont {I.}~\bibnamefont {Medina}}, \bibinfo {author} {\bibfnamefont {E.}~\bibnamefont {Cruz}}, \bibinfo {author} {\bibfnamefont {Q.}~\bibnamefont {Huang}},\ and\ \bibinfo {author} {\bibfnamefont {C.~B.}\ \bibnamefont {Mendl}},\ }\bibfield  {title} {\bibinfo {title} {Simulating quantum circuits using tree tensor networks},\ }\href {https://doi.org/10.22331/q-2023-03-30-964} {\bibfield  {journal} {\bibinfo  {journal} {Quantum}\ }\textbf {\bibinfo {volume} {7}},\ \bibinfo {pages} {964} (\bibinfo {year} {2023})}\BibitemShut {NoStop}%
\bibitem [{\citenamefont {Sozykin}\ \emph {et~al.}(2022)\citenamefont {Sozykin}, \citenamefont {Chertkov}, \citenamefont {Schutski}, \citenamefont {Phan}, \citenamefont {Cichocki},\ and\ \citenamefont {Oseledets}}]{TTOPT}%
  \BibitemOpen
  \bibfield  {author} {\bibinfo {author} {\bibfnamefont {K.}~\bibnamefont {Sozykin}}, \bibinfo {author} {\bibfnamefont {A.}~\bibnamefont {Chertkov}}, \bibinfo {author} {\bibfnamefont {R.}~\bibnamefont {Schutski}}, \bibinfo {author} {\bibfnamefont {A.-H.}\ \bibnamefont {Phan}}, \bibinfo {author} {\bibfnamefont {A.}~\bibnamefont {Cichocki}},\ and\ \bibinfo {author} {\bibfnamefont {I.}~\bibnamefont {Oseledets}},\ }\href {https://arxiv.org/abs/2205.00293} {\bibinfo {title} {Ttopt: A maximum volume quantized tensor train-based optimization and its application to reinforcement learning}} (\bibinfo {year} {2022}),\ \Eprint {https://arxiv.org/abs/2205.00293} {arXiv:2205.00293 [cs.LG]} \BibitemShut {NoStop}%
\bibitem [{\citenamefont {Hao}\ \emph {et~al.}(2022)\citenamefont {Hao}, \citenamefont {Huang}, \citenamefont {Jia},\ and\ \citenamefont {Peng}}]{Optimiz}%
  \BibitemOpen
  \bibfield  {author} {\bibinfo {author} {\bibfnamefont {T.}~\bibnamefont {Hao}}, \bibinfo {author} {\bibfnamefont {X.}~\bibnamefont {Huang}}, \bibinfo {author} {\bibfnamefont {C.}~\bibnamefont {Jia}},\ and\ \bibinfo {author} {\bibfnamefont {C.}~\bibnamefont {Peng}},\ }\bibfield  {title} {\bibinfo {title} {A quantum-inspired tensor network algorithm for constrained combinatorial optimization problems},\ }\bibfield  {journal} {\bibinfo  {journal} {Frontiers in Physics}\ }\textbf {\bibinfo {volume} {10}},\ \href {https://doi.org/10.3389/fphy.2022.906590} {10.3389/fphy.2022.906590} (\bibinfo {year} {2022})\BibitemShut {NoStop}%
\bibitem [{\citenamefont {Fang}\ and\ \citenamefont {Lele}(2022)}]{spiking_qubo}%
  \BibitemOpen
  \bibfield  {author} {\bibinfo {author} {\bibfnamefont {Y.}~\bibnamefont {Fang}}\ and\ \bibinfo {author} {\bibfnamefont {A.~S.}\ \bibnamefont {Lele}},\ }\bibfield  {title} {\bibinfo {title} {Solving quadratic unconstrained binary optimization with collaborative spiking neural networks},\ }in\ \href {https://doi.org/10.1109/ICRC57508.2022.00021} {\emph {\bibinfo {booktitle} {2022 IEEE International Conference on Rebooting Computing (ICRC)}}}\ (\bibinfo {year} {2022})\ pp.\ \bibinfo {pages} {84--88}\BibitemShut {NoStop}%
\bibitem [{\citenamefont {Nakatani}(2018)}]{dmrg}%
  \BibitemOpen
  \bibfield  {author} {\bibinfo {author} {\bibfnamefont {N.}~\bibnamefont {Nakatani}},\ }\bibfield  {title} {\bibinfo {title} {Matrix product states and density matrix renormalization group algorithm},\ }in\ \href {https://doi.org/https://doi.org/10.1016/B978-0-12-409547-2.11473-8} {\emph {\bibinfo {booktitle} {Reference Module in Chemistry, Molecular Sciences and Chemical Engineering}}}\ (\bibinfo  {publisher} {Elsevier},\ \bibinfo {year} {2018})\BibitemShut {NoStop}%
\bibitem [{\citenamefont {Ali}(2025)}]{melocoton}%
  \BibitemOpen
  \bibfield  {author} {\bibinfo {author} {\bibfnamefont {A.~M.}\ \bibnamefont {Ali}},\ }\href {https://arxiv.org/abs/2502.05981} {\bibinfo {title} {Explicit solution equation for every combinatorial problem via tensor networks: Melocoton}} (\bibinfo {year} {2025}),\ \Eprint {https://arxiv.org/abs/2502.05981} {arXiv:2502.05981 [cs.ET]} \BibitemShut {NoStop}%
\bibitem [{\citenamefont {Ali}\ \emph {et~al.}(2023)\citenamefont {Ali}, \citenamefont {Delgado}, \citenamefont {Roura},\ and\ \citenamefont {de~Leceta}}]{Initi}%
  \BibitemOpen
  \bibfield  {author} {\bibinfo {author} {\bibfnamefont {A.~M.}\ \bibnamefont {Ali}}, \bibinfo {author} {\bibfnamefont {I.~P.}\ \bibnamefont {Delgado}}, \bibinfo {author} {\bibfnamefont {M.~R.}\ \bibnamefont {Roura}},\ and\ \bibinfo {author} {\bibfnamefont {A.~M.~F.}\ \bibnamefont {de~Leceta}},\ }\href@noop {} {\bibinfo {title} {Efficient finite initialization for tensorized neural networks}} (\bibinfo {year} {2023}),\ \Eprint {https://arxiv.org/abs/2309.06577} {arXiv:2309.06577 [cs.LG]} \BibitemShut {NoStop}%
\bibitem [{\citenamefont {Perron}\ and\ \citenamefont {Furnon}()}]{ortools}%
  \BibitemOpen
  \bibfield  {author} {\bibinfo {author} {\bibfnamefont {L.}~\bibnamefont {Perron}}\ and\ \bibinfo {author} {\bibfnamefont {V.}~\bibnamefont {Furnon}},\ }\href {https://developers.google.com/optimization/} {\bibinfo {title} {Or-tools}}\BibitemShut {NoStop}%
\bibitem [{\citenamefont {{D-Wave Systems Inc.}}()}]{dimod}%
  \BibitemOpen
  \bibfield  {author} {\bibinfo {author} {\bibnamefont {{D-Wave Systems Inc.}}},\ }\href {https://docs.dwavesys.com/docs/latest/doc_dimod.html} {\bibinfo {title} {dimod}}\BibitemShut {NoStop}%
\end{thebibliography}%

\end{document}